\documentclass[12pt]{article}
\usepackage{epsfig}
\newcommand{\D}{/\! \! \! \! D}
\begin{document}
\thispagestyle{empty}
\begin{flushright}
MCTP-03-15\\
%(Version of~\today)\\
\end{flushright}
\vspace{0.5cm}
\begin{center} 
{\Large \bf 
The Large-$x$ Factorization of the Longitudinal Structure Function}\\[3mm]
\vspace{1.7cm}
{\sc \bf R. Akhoury and M. G. Sotiropoulos}\\[1cm]
\begin{center} \em 
Michigan Center for Theoretical Physics\\
Randall Laboratory of Physics\\ 
University of Michigan, Ann Arbor, MI 48109 
\end{center}\end{center}
\vspace{3cm}
\begin{abstract}
A leading-twist factorization formula is derived for the longitudinal 
structure function in the $x\rightarrow 1$ limit of 
deeply inelastic scattering.  
This is achieved by defining a new jet function which is gauge independent 
and probes the transverse momentum of the struck parton in the target.  
In moment space, terms of order $\ln^k N/N$, which are 
the leading ones for $F_L$, are shown to be resummable through the cusp 
anomalous dimension $\gamma_K$ and the anomalous dimension 
$\gamma_{J^\prime}$ of the new jet function. 
This anomalous dimension is computed to ${\cal O}(\alpha_s)$. 
The suggested factorization for $F_L$ reproduces the fixed order results 
known to ${\cal O}(\alpha_s^2)$.
The general ideas for resumming the terms of order $\ln^k N/N$ in 
moment space may be extended to the other structure functions and to 
other inclusive processes near the elastic limit.
\end{abstract}

\vspace*{\fill}
% footnotes ------------------------------------
\noindent { e-mail: akhoury@umich.edu}
% main text ------------------------------------

\newpage
\section{Introduction}
Typical applications of QCD perturbation theory involve processes with two 
mass scales, the momentum transfer $Q$ and the scale of strong interactions 
$\Lambda_{\rm QCD}$ with $Q^2 \gg \Lambda_{\rm QCD}^2$.
Within this large family of hard processes a particular class is of great 
theoretical and phenomenological importance. 
These are the quasi-elastic processes, characterized by a third scale $M^2$,
the invariant mass of the final states, with $Q^2 \gg M^2$. 
Here perturbation theory can be used to study the transition from 
short-distance dynamics ($M^2 \gg \Lambda^2_{\rm QCD}$) to long-distance 
dynamics as the elastic limit is approached ($M^2 \rightarrow 
\Lambda^2_{\rm QCD}$). Deeply inelastic scattering (DIS) structure functions 
at large Bjorken-$x$, Drell-Yan production near the partonic threshold and 
thrust in $e^+$-$e^-$ annihilation near the two-jet limit are all 
characteristic examples of quasi-elastic processes. 
In this paper we focus on the large-$x$ DIS structure functions for 
definiteness, but we expect our discussion to be generalizable to all 
quasi-elastic processes.

Near the elastic phase-space region $x \rightarrow 1$ of DIS the 
coefficient functions receive corrections in $1-x$.
For $F_2$ the leading power in $x$ corrections are terms of the form 
$\alpha_s^n [\ln^m(1-x)/(1-x)]_+$, with $m \le 2n-1$. The formalism of 
resumming this class of corrections has been developed in 
Refs.~\cite{Sterman,CatTre,KorMar} and recently reviewed in 
Ref.~\cite{CoLaSte}.
A key step is the factorization of the perturbation theory diagrams at large
$x$ in terms of a hard amplitude, a universal soft function, 
and a jet function. 
We call this factorization "light-cone expansion", distinct from the 
Wilson operator product expansion.
The soft function and the jet functions contain singular pieces like 
$\ln^k(1-x)/ (1-x)_+$ and $1/ (1-x)_+$ respectively, which are resummed. 
All these corrections correspond to ${\cal O}(\alpha_s^n \ln^k N)$ terms 
in moment space. 

The question we consider here is whether a formalism beyond the 
light-cone expansion can be constructed  to resum terms that do 
not correspond to power singularities in $x$ 
and are still leading power in $Q$.  
A class of such terms are the simple logarithmic corrections of order 
$\alpha_s^n \ln^m(1-x)$, which in moment space generate 
${\cal O}(\alpha_s^n \ln^k N /N)$ power suppressed contributions. 
The cleanest context where such corrections are encountered is in the 
longitudinal structure function $F_L$. 
$F_L$, as distinct from $F_2$, does not contain any singular distributions 
in $x$. Its leading power in $x$ corrections are all logarithmic and they 
start at order $\alpha_s^2$. 
A phenomenological motivation for studying the resummation of $F_L$ 
near the elastic limit comes from considering the ratio $R$ of the 
longitudinal and the transverse cross 
sections :
\begin{equation}
R~=~{\sigma_L \over \sigma_T}
~=~{F_L \over (1+{4 M^2 x^2 \over Q^2})F_2-F_L}, 
\label{R}
\end{equation}
where $F_i$, $i=2,L$ denote the structure functions, $M$ is the target mass 
and $x=-q^2 /( 2p\cdot q)$. The longitudinal structure function $F_L$ vanishes
due to helicity conservation in the naive parton model and thus $R$ provides 
us with a clean test of perturbative QCD. 
There already exist some experimental data \cite{flexpt} on this quantity 
and more is expected in the near future.
Leading ${\cal O}(\alpha_s)$ and next to leading ${\cal O}(\alpha_s^2)$ 
perturbative corrections to $F_L$ have been calculated 
\cite{DDKS}-\cite{ZvN} including the target mass corrections. 
Comparison with the data indicates that the theoretical
predictions lie systematically below the experimental points, especially 
in the large $x$ region. 
This may be indicative of rather large higher twist corrections
\cite{phenfit}.
In order to shed more light on this issue it is worth studying the effects 
of resummation at the leading twist level.

The resummability of quasi-elastic corrections that do not correspond to 
singular distributions was considered in Ref.~\cite{ASS1}. 
There it was argued that final state power suppressed corrections 
of order $\alpha_s^n (1-x)^p \ln^k(1-x)$, with $p \ge 0$, can be 
identified and resummed through an expansion of the 
final state interactions in terms of higher dimensional bilocal operators. 
The expectation values of these operators define classes of new jet 
functions that generalize the jet function of the light-cone expansion.
In this new expansion operators of different dimensionality all 
contribute to the leading power in $Q$ (leading twist)
after integration over the final state phase-space. 
However, higher dimensional operators produce terms suppressed by higher 
powers of $1-x$ near the elastic limit $x \rightarrow 1$.  
In Ref.~\cite{ASS1} rules were given for identifying independent sets 
of these higher dimensional operators. Once they are constructed, their 
anomalous dimensions can be used to resum the power suppressed logarithmic 
corrections, just like in the case of the light cone expansion.

Within the context of the above formal construction, in this work we consider 
the longitudinal structure function in the elastic limit as a specific 
example of a quasi-elastic process. 
We shall show that by analyzing the perturbative diagrams in the 
appropriate infrared limits we, indeed, obtain the operator expansion 
suggested in Ref.~\cite{ASS1}. 
We accomplish this not by dimensional arguments but through a diagrammatic 
analysis by explicitly considering the subleading terms in the eikonal 
approximation for the final state. 
The result is a factorization formula for $F_L$ that is used to resum 
in moment space all the ${\cal O}(\ln^kN/N)$ corrections.
We find it remarkable that such power suppressed corrections can be 
treated systematically using the standard methods for analyzing 
the infared region of perturbation theory. Simple logarithms of 
$1-x$ come from the pinch singular surfaces of Feynman diagrams. 
This observation, we believe has not been considered in the literature 
so far.

The general ideas for resumming terms of order $\ln^kN/N$ in moment space
can be extended to other structure functions and to other inclusive processes
near the elastic limit. For example, it has been noticed \cite{laenen} in the
context of Higgs production that such corrections can be phenomenologically
important. They may also be of relevance in renormalon phenomenology
\cite{renormalon}. Thus, the analysis of the resummation of the 
${\cal O}(\ln^kN/N)$ corrections is easiest done for the case of the
longitudinal structure function, but as discussed, the applications 
go well beyond that to phenomenologically more interesting situations
which we hope to address in the future.

The structure of this paper is as follows.
In section 2 we set notations and review known fixed order results.
In section 3 we briefly summarize the light-cone expansion for $F_2$, 
which is used as a basis for what follows. 
Then we go on to show how the new jet function, characteristic of $F_L$, 
arises from the infrared singularities of Feynman diagrams, thus explicitly 
constructing the relevant jet operator of Ref.~\cite{ASS1}. 
Here we also study the renormalization group properties of the 
functions entering the $F_L$ factorization formula.
In section 4 we present an explicit one-loop calculation of the anomalous 
dimension of the new jet function, which is subsequently used in the 
$F_L$ resummation formula. 
The resummed coefficient functions are found in agreement with the 
fixed order results. 
We summarize our conclusions in the final section. 
Certain useful formulae are listed in the Appendix.

\section{Notations and known results}
\setcounter{equation}{0}

We consider the DIS structure functions $F_2, F_L$ of a massless quark of 
electric charge $Q_f=1$ and with collinear divergences dimensionally 
regularized in $D=4-2\epsilon$ dimensions. 
Their perturbative expansion is 
\begin{equation}
F_{\rm r}(x, Q^2,\epsilon) = 
\sum_{n=1}^{\infty} \left( \frac{\alpha_s}{\pi} \right)^n 
F_{\rm r}^{(n)}(x, Q^2,\epsilon) \, , 
\ \ \ {\rm r} = 2, L \, .
\label{alphaexp}
\end{equation} 
The DIS quark tensor for an electromagnetic probe is defined as 
\begin{equation}
W_{\mu \nu}(p, q) =  \frac{1}{4 \pi}
\int d^Dy \; {\rm e}^{-iq\cdot y} \; 
\langle p | j_\mu^\dagger(0) \, j_\nu(y) |p \rangle\ 
= \frac{1}{4 \pi}  \int d PS_n 
\langle {\cal M}^*_R {\cal M}_L \rangle \, , 
\label{wtens}
\end{equation}
where $PS_n$ is the Lorenz invariant phase space for $n$ particles in the 
final state and ${\cal M}_{L, R}$ are the left and right production  
amplitudes.  
Structure functions are obtained via the projections 
$F_{\rm r} = P^{\mu \nu}_{\rm r} \, W_{\mu \nu}$, for ${\rm r} = 2, L$, 
given at $D=4$ by 
\begin{equation}
P_L^{\mu \nu} = \frac{8 x^2}{Q^2} p^\mu p^\nu , 
\ \ \ \ \
P_2^{\mu \nu} = -\eta^{\mu\nu} +\frac{3}{2} P^{\mu\nu}_L. 
\label{projectors}
\end{equation} 
The normalization in Eq.~(\ref{wtens}) is chosen so that 
$F_2^{(0)} = \delta(1-x)$.

The light cone is defined by the two dimensionless vectors $v^\mu$ and 
$\bar{v}^\mu$ normalized as $v\cdot \bar{v} =1$ and we fix the  $v^\mu$ (+) 
direction to be parallel to the incoming quark momentum $p$. 
Then the $\bar{v}^\mu$ ($-$) direction is set by the vector 
\mbox{$\bar{q}=x p+q$}. 
We denote by $p_f$ the total final state momentum, $p_f=p+q$. 
The light cone (Sudakov) decomposition of the external momenta is      
\begin{equation}
p^\mu=\frac{Q}{\sqrt{2}} v^\mu \, ,  \  \  \
p_f= (1-x) \frac{Q}{\sqrt{2}} v^\mu 
+\frac{Q}{\sqrt{2}x} \bar{v}^\mu \, , \ \ \ 
\bar{q} = \frac{Q}{\sqrt{2}x} \bar{v}^\mu,  
\label{vdef}
\end{equation}
and for every momentum vector $k^\mu$  
\begin{equation} 
k^\mu= (\bar{v}\cdot k) v^\mu + (v\cdot k) \bar{v}^\mu + k_\perp^\mu \, ,  
\ \  \
d^Dk =d (\bar{v}\cdot k)\,  d(v\cdot k) \,  d^{D-2} k_\perp  .
\label{lcdec}
\end{equation}
The two massless particle phase space in this basis and for azimuthically 
symmetric integrands becomes 
\begin{equation}
\int d PS_2(k_1, k_2) = \frac{1}{8\pi}\frac{1}{\Gamma(1-\epsilon)} 
\left(\frac{4\pi x}{(1-x) Q^2}\right)^\epsilon
\int_0^1 d \beta_2 \, \beta_2^{-\epsilon}(1-\beta_2)^{-\epsilon},
\label{PS2fin}
\end{equation}  
with the Sudakov variable $\beta_2$  defined as 
$\beta_2=(v\cdot k_2)/(v\cdot \bar{q})$. 
The three massless particle phase space in terms of Sudakov variables is 
given in the appendix. 

At high momentum transfer, $Q^2 \gg \Lambda_{\rm QCD}^2$, and fixed $x$ 
all structure functions satisfy the Wilson operator product expansion. 
At leading twist this can be conveniently written as a factorization theorem 
in terms of the convolution symbol 
\begin{equation}
(h\otimes f)(x) = \int_0^1 dx_1  \int_0^1 dx_2 \, \delta(x-x_1 x_2) \,
h(x_1)\, f(x_2). 
\label{convsym}
\end{equation}
Specifically for $F_L$ we have 
\begin{equation}
F_L(x, Q^2,\epsilon) = \bar{C}_L(x,Q^2/\mu^2,\alpha_s(\mu^2)) \otimes 
f(x,\epsilon,\alpha_s(\mu^2)) +{\cal O}(\Lambda_{\rm QCD}^2/Q^2), 
\label{OPEfac}
\end{equation} 
with $\mu$ the factorization scale and $f$ the quark distribution function 
of the quark target. The coefficient functions are determined given a 
factorization scheme, which will be the $\overline{\rm MS}$ scheme for 
this paper. The coefficient function $\bar{C}_L$ has been calculated 
to ${\cal O}(\alpha_s^2)$ and the results are \cite{SG,ZvN,Zijlstra} 
\begin{equation}
\bar{C}_L^{(1)}(x,Q^2/\mu^2) = C_F x, 
\label{Cl1} 
\end{equation}
\begin{equation}
\bar{C}_L^{(2)}(x,Q^2/\mu^2) =  \left[
\frac{1}{2} \left(P_{qq}^{(1)} \otimes \bar{C}_L^{(1)}\right)(x) 
-\frac{1}{4}\beta_1 \, 
\bar{C}_L^{(1)}(x) \right] 
\ln \frac{Q^2}{\mu^2} +\bar{C}_L^{(2)}(x,1),  
\label{Cl2}
\end{equation}
with the well known functions  
\begin{equation}
P_{qq}^{(1)}=C_F \frac{1+x^2}{(1-x)_+} +C_F\frac{3}{2} \delta(1-x), 
\ \ \ \ \ \ \beta_1=\frac{11}{3}C_A-\frac{2}{3}N_f, 
\label{split} 
\end{equation}
and
\begin{eqnarray} 
\bar{C}_L^{(2)}(x,1) &=& 
\frac{1}{2} C_F^2\ln^2(1-x) + 
\left[\frac{9}{4} -2 \zeta(2)\right]C_F^2 \ln (1-x) 
\nonumber \\ 
& \ & 
+ \left[ (\zeta(2)-1)C_F C_A -\frac{1}{4} \beta_1 C_F \right] \ln(1-x) 
+{\rm reg}. 
\label{cl2}
\end{eqnarray}
In the above expression only terms that are singular in the 
$x\rightarrow 1$ limit have been retained. It is these logarithms of 
$1-x$ that are the focus of the present study.

\section{The factorization formula for $F_L$ }
\setcounter{equation}{0}

We shall show that $F_L$ satisfies a novel factorization formula in the 
$x\rightarrow 1$ limit which can be used to sum the singular $\ln^k(1-x)$ 
contributions in the coefficient function, (see Eq.~(\ref{cl2})). 
This  factorization formula involves a new jet function that contains these 
leading corrections to $F_L$ in the large $x$ limit, which in moment space 
are of ${\cal O}(\ln^k N/N)$, $N$ being the moment variable. 
Such contributions are also present in $F_2$, although they are subleading
to the ones that are of ${\cal O}(\ln^k N)$ in moment space. 

\subsection{The new jet function} 

Before presenting the factorization formula for $F_L$ in the 
$x\rightarrow 1$ limit it is worth reviewing the light-cone expansion 
for $F_2$, since the latter will serve as a prototype for the former. 
The leading power in $x$ factorization of $F_2$ can be cast in the following 
form \cite{Sterman}
\begin{eqnarray}
F_2(x, Q^2) &=&  |H_2(Q^2)|^2 \, 
\int_x^1 dx' \int_0^{x'-x} dw\; J\left((x'-x-w)Q^2\right)\; V(w) \; \phi(x')
\nonumber\\
&\equiv& 
|H_2(Q^2)|^2 J \otimes V \otimes \phi\, . 
\label{f2fac}
\end{eqnarray}
$H_2(Q^2)$ is the hard scattering function on either side of the 
final state cut. $|H_2(Q^2)|^2$ is by construction UV dominated, i.e. 
all its internal lines are far from their mass shell. 
$V$ is the universal soft radiation function defined as
\begin{eqnarray}
V(w) &=& \int dy^- {\rm e}^{-iw y^- p \cdot \bar v} \;
\langle 0| \Phi^\dagger_v(0, -\infty) \; 
\Phi_{\bar{v}}(0, y^- \bar v) \;  
\Phi_{v}(y^- \bar v, -\infty) |0 \rangle
\nonumber \\ 
&\equiv & {\rm FT}^{(1)}_{-wp} \; 
\langle 0| \Phi^\dagger_v(0, -\infty) \; 
\Phi_{\bar{v}}(0, y^- \bar v) \;  
\Phi_{v}(y^- \bar v, -\infty) |0 \rangle \, .
\label{Vdef}
\end{eqnarray}
${\rm FT}^{(1)}_{-wp}$ stands for the Fourier transformation along the 
light-cone direction $-v$.
The Wilson line operator starting in the far light-cone past and defined
along 
the direction $v^\mu$ is given by  
\begin{equation}
\Phi_v(x+tv, -\infty) = 
P \exp\left( -i g_s \int_{-\infty}^t d s \, 
v^\mu \, A_\mu( x+s v ) \right) \, .
\label{Phidef}
\end{equation} 
Its perturbative expansion generates the Feynman rules for the eikonal 
approximation.
$\phi$ is the light-cone parton distribution function.
Its operator definition, for the scattering of quarks, is 
\begin{equation}
\phi(x') =
{\rm FT}^{(1)}_{x'p}\, \langle p| \psi(0) \not{\bar{v}} \; 
\Phi_{\bar{v}}(0,y^-) \, \bar{\psi}(y)|p\rangle \otimes V^{-1}\, , 
\label{phidef}
\end{equation} 
This distribution function is not identical to $f$ used in 
Eq.~(\ref{OPEfac}), but they both share the feature that the only 
singularities they contain come from collinear propagation along the 
initial ($v$) light cone direction. 
The role of $V^{-1}$ in Eq.~(\ref{phidef}) is to remove soft contributions,
which would otherwise be double-counted in both $\phi$ and $V$. 

Finally, $J$ is the current jet function that describes the stream of almost 
collinear particles with invariant mass  $(x'-x-w)Q^2$ moving along 
the $\bar{v}$ light-cone direction. 
$J$ is defined as a gauge invariant and soft subtracted fermion propagator  
\begin{equation}
J\left((1-z)Q^2\right)= {\rm FT}^{(4)}_{q+zp}\;  \langle 0|  
\Phi^\dagger_v(0, -\infty) \; \psi(0) \, 
\bar{\psi}(y) \;  \Phi_{v}(y, -\infty) |0 \rangle \otimes V^{-1},
\label{jdef}
\end{equation}
where ${\rm FT}^{(4)}_{q+zp}$ denotes the four-dimensional Fourier 
transformation with respect to the total jet momentum $q+zp$.

We note that the factorization theorem of Eq.~(\ref{f2fac}) is written 
in terms of UV renormalized effective non-local operators. 
When one analyzes the contributions to the characteristic 
functions $|H_2|^2$, $V$, $J$ and $\phi$ coming from classes 
of graphs of the full theory, it is often the case that 
the assignment to soft and jet functions of these contributions is 
gauge dependent with gauge independence occuring only for their convolution, 
i.e. the final result. 
Here, however, the characteristic functions are defined in terms of gauge 
invariant operators, thus we can classify singularities and enhancements 
as soft or collinear in an unambiguous way. The cost for this is the need 
to perform infrared subtractions, as denoted for instance by the denominator 
$V$ in Eq.~(\ref{jdef}) for the jet function. 

In Ref.~\cite{ASS1} it was argued that $F_L$ satisfies the following 
factorization formula:
\begin{equation}
F_L(x, Q^2, \epsilon) = |H_L(Q^2)|^2 J^\prime \otimes V \otimes \phi \, .
\label{flfac}
\end{equation} 
The factors $V$ and $\phi$ are the same as in the $F_2$ case, 
see Eqs.\ (\ref{Vdef}), (\ref{phidef}). 
The new jet function $J^\prime$, is defined as the expectation value 
of a bi-local operator that probes the transverse momentum of the 
outgoing struck quark. Explicitly,
\begin{equation}
J^\prime((1-z) Q^2) = 
 \left(\frac{1}{4\pi} \; \frac{8 x^2}{Q^2} \right) \; 
 {\rm FT}^{(4)}_{q+zp} \;  
\langle 0| \Phi_{v}^\dagger (0, -\infty) \; \D_\perp \psi(0) \; 
\D_\perp\bar{\psi}(y) \; \Phi_v (y, -\infty) |0 \rangle \otimes V^{-1}.
\label{Jprimedef}
\end{equation}
The conventional normalization factors of 
Eqs.~(\ref{wtens}), (\ref{projectors}) have been included in this 
definition.  
The operator defining $J^\prime$ is the only gauge invariant bi-local 
operator that is non-vanishing by the massless quark equations of motion 
or the $F_L$ projection of Eq.~(\ref{projectors}), 
and although it is of higher dimension than the corresponding one that 
defines $J$, Eq.~(\ref{jdef}), it contributes to leading twist. 
All the terms that it generates are of order $(1-x)^0$, i.e. only logarithms. 
It is therefore the appropriate operator for resumming the leading
logarithmic 
quasi-elastic corrections in $F_L$. 
In Ref.~\cite{ASS1} the new jet operator was identified by requiring it 
satisfied all the above properties. In the next subsection we show that 
it actually comes out naturally from the infrared properties of the 
Feynman diagrams that contribute to $F_L$ at leading twist. 

The jet function $J^{\prime}$ is depicted in fig.~1(a), and  
the factorization expression for $F_L$ is depicted in fig.~1(b). 
We emphasize that the factor $V\otimes \phi$ is common to every 
DIS observable, and that the only difference between $F_2$ and $F_L$ at 
their respective leading power in $x$ level is that 
they are different Lorentz projections of the same tensor $W_{\mu\nu}(p,q)$, 
hence they differ in the jet functions entering their respective 
factorization formulae.

\subsection{Operators beyond the eikonal approximation} 

In order to prove the factorization formula in Eq.~(\ref{flfac}) in this
subsection  we  follow the constructive approach. 
We analyze the perturbation theory diagrams 
and show how the operator in Eq.~(\ref{Jprimedef}) emerges once one goes 
beyond the eikonal approximation for $F_L$. 
Infrared singularities and ensuing corrections
arise from characterisitc regions in momentum space called 
pinch surfaces. These are the same for all observables, therefore the 
classification of internal lines in an $F_L$ diagram are the same as in 
any other observable, i.e. we have collinear, soft and ultraviolet lines.
The factorization of the soft lines from the incoming $v$-jet is as in $F_2$. 
We, therefore, concentrate on the factorization of $\bar{v}$-collinear lines 
from the initial state $v$-jet since this is where the difference between 
$F_L$ and $F_2$ lies.    

Consider the configuration shown in fig.~2, in which the depicted gluon 
belongs to the $\bar{v}$-jet. Clearly the fermion line carrying momentum 
$(p-l)$ is far off-shell and belongs to the hard scattering subdiagram in 
the notation of reduced diagrams. If, however, we were to factorize the 
collinear gluon as usual by eikonalizing the fermion propagator then 
this, in general leading, IR contribution would be projected out in 
$F_L$ because of the contraction with $p^\mu$ at the current vertex. 
Note that also in the case where the gluon is soft, the leading eikonal 
contribution is similarly projected out. Therefore, we must go beyond the 
eikonal approximation.  
 
The graph in fig.~2 is proportional to 
\begin{equation}
g_s\frac{1}{- 2 p \cdot l + l^2} \not{p} (\not{p} - \not{l}) \gamma^\lambda  
\approx g_s\frac{1}{2 p \cdot l} \not{p} \not{l}\gamma^\lambda 
=g_s\left( \gamma^\lambda-\frac{v^\lambda}{v\cdot l} \not{l} \right)
+g_s \frac{1}{2v\cdot l}\not{l}\gamma^\lambda\not{v}.
\label{fig1}
\end{equation}
We define the transverse projector 
$\omega^\alpha_\beta =\delta^\alpha_{\perp\beta}$ and decompose every 
$\gamma$ matrix in light cone components as
\begin{equation}
\gamma^\lambda=\bar{v}^\lambda \not{v}
+v^\lambda \not{\bar{v}}+(\omega \gamma)^\lambda. 
\label{gammadec}
\end{equation}
We also define the effective vertex 
\begin{equation}
O^\lambda(l) =g_s\left( (\omega\gamma)^\lambda-(l\omega\gamma)
\frac{v^\lambda}{v\cdot l}\right). 
\label{Odef}
\end{equation}
It enjoys the following properties
\begin{equation}
O^\lambda(l) l_\lambda=0, \ \ \  O^\lambda(l) v_\lambda = 0, \ \ \ 
O^\lambda(l) \bar{v}_\lambda=-g_s(l\omega\gamma)\frac{1}{v\cdot l}, 
\ \ \ O^\lambda(l) O_\lambda(l^\prime)=-2g_s^2,  
\label{Oprop1}
\end{equation}
and
\begin{equation}
O^\lambda(l) ... O_\lambda(l^\prime) = g_s^2
(\omega \gamma)^\lambda ... (\omega \gamma)_\lambda. 
\label{Oprop2}
\end{equation}
Then the factor in Eq.~(\ref{fig1}) becomes 
\begin{equation}
O^\lambda(l) -\frac{1}{2}O^\lambda(l)\not{\bar{v}}\not{v} 
+g_s \bar{v}^\lambda\not{v} 
+g_s \frac{1}{2 v\cdot l}(l\omega\gamma)(\omega\gamma)^\lambda \not{v}.
\label{fig1next}
\end{equation}
If only one gluon from the $\bar{v}$-jet is connected to the hard scattering, 
as in fig.~2, then only the first term in (\ref{fig1next}) survives 
the massless Dirac equation. Recall that the leading spin structure of the 
$v$-jet is $\not{v}$.  
We observe that the vertex $O^\lambda(l)$ is indeed generated 
by the ${\cal O}(g_s)$ expansion of $\Phi_v(0, -\infty) \, \D_\perp \psi(0)$, 
which is the part of the operator in Eq.~(\ref{Jprimedef}) that sits to 
the left of the final state cut.
% The ${\cal O}(g_s^0)$ expansion of the operator yields terms 
% proportional to $\bar{p}_\perp$ which vanish for fig.~2 due to the final 
% state kinematics, i.e. in the Breit frame the $\bar{v}$-jet has  
% no transverse momentum.

To show that the operator in Eq.~(\ref{Jprimedef}) gives the leading 
fragmentation contributions for $F_L$  to all orders we proceed inductively 
by considering the coupling of two $\bar{v}$-jet gluons to the 
hard scattering, fig.~3. 
Because the hard scattering subgraph is UV dominated, its Dirac structure 
is again $\gamma^\mu$ hence for $F_L$ the same $/ \! \! \! p$ factor enters  
at the current vertex. Then the graph of fig.~2 is proportional to 
\begin{eqnarray}
&\quad&
(T_1 T_2) \left\{ O^{\lambda_1}(l_1+l_2) 
\left[-g_s\frac{1}{2\, p\cdot l} (\not{p}-\not{l}_2) \gamma^{\lambda_2} 
\right] 
\right.
\nonumber \\
&\quad& 
+\frac{1}{2\, \bar{v}\cdot p} \not{\bar{v}} O^{\lambda_1}(l_1+l_2) 
O^{\lambda_2}(l_2) +g_s\frac{\bar{v}^{\lambda_1}}{\bar{v}\cdot p} 
O^{\lambda_2}(l_2) 
\nonumber \\
&\quad& \left.
+ g_s\frac{1}{2 (v\cdot (l_1+l_2))(\bar{v}\cdot p)} 
\left[(l_1+l_2)\omega\gamma\right](\omega \gamma)^{\lambda_1} 
O^{\lambda_2}(l_2) \right\} +(1\leftrightarrow 2),
\label{fig2}
\end{eqnarray}
where terms that will vanish by the massless Dirac equation have been
dropped. 
There is a hierarchy among the terms of the above expression once this factor 
is imbedded into a cross section diagram. The leading contribution comes 
from the first term and with its factor in square brackets eikonalized, 
i.e. from 
\begin{eqnarray}
&\ &
(T_1 T_2) O^{\lambda_1}(l_1+l_2) 
\left(-g_s\frac{v^{\lambda_2}}{v\cdot l_2} \right) +(1 \leftrightarrow 2)=
\nonumber \\
&\ &
-g_s^2 (T_1T_2)\left( (\omega\gamma)^{\lambda_1}
\frac{v^{\lambda_2}}{v\cdot l} 
-[(l_1+l_2)\omega\gamma]\frac{v^{\lambda_1}}{v\cdot (l_1+l_2)}
\frac{v^{\lambda_2}}{v\cdot l_2}\right) +(1\leftrightarrow2).
\label{fig2lead}
\end{eqnarray} 
All the other terms in Eq.~(\ref{fig2}) are suppressed by at least one power 
of $(1-x)$ in the cross section, relative to the contribution that 
has been retained.
We observe that Eq.~(\ref{fig2lead}) is indeed the vertex in momentum space
generated  by the ${\cal O}(g_s^2)$ expansion  of 
$\Phi_v(0,-\infty) \, \D_\perp \psi(0)$.  
More specifically, the first term corresponds to the piece that is 
${\cal O}(g_s)$ in the Wilson line and  ${\cal O}(g_s)$ in the covariant 
derivative and the second term corresponds to the piece that is 
${\cal O}(g_s^2)$ in the Wilson line and ${\cal O}(g_s^0)$ in the 
covariant derivative.

The above argument can be generalized straightforwardly for an arbitrary 
number of $\bar{v}$-jet gluons attached to the hard scattering to the left 
of the final state cut. The leading contribution at $x \rightarrow 1$ is 
generated by $\Phi_v(0,-\infty) \, \D_\perp \psi(0)$. This is how we show 
constructively, i.e. from analyzing the diagrams, that the new jet function 
$J^\prime$, as defined by the operator of Eq.~(\ref{Jprimedef}), will contain 
the leading, ${\cal O}( (1-x)^0)$, terms in the longitudinal cross section.  
Note that the operator in Eq.~(\ref{Jprimedef}) contributes at leading 
twist level. It is also the only operator that generates ${\cal O}((1-x)^0)$ 
contributions from the fragmentation of the final state jet. 
It can be shown that any other operator with higher number of covariant 
derivatives than in $J^\prime$ will genarate contributions of 
${\cal O}((1-x)^k)$, $k=1,2,...$
 
Once the jet function $J^\prime$ is defined in terms of the expectation 
value of an operator it can be calculated in perturbation theory.  
However, the momentum integrals that are obtain in this calculation
will also contain contributions in which the gluons coupling to the 
$\bar{v}$-jet are soft. It is a well known (and non-trivial) statement that 
such contributions can be expressed in terms of expectation values of Wilson
lines. Such contributions are removed by the denominator in 
Eq.~(\ref{Jprimedef}) so that $J^\prime$ is by definition free from soft 
contributions. The latter ones will be contained only in the function $V$, 
Eq.~(\ref{Vdef}).  
In fixed order perturbation theory this amounts to performing IR subtractions 
from the expressions generated by the numerator in order to obtain a 
$J^\prime$ jet function that contains only enhancements from final state 
fragmentation. 

There is also one more subtlety that arises when dealing with the function 
$J^\prime$, which is again characteristic of $F_L$. The UV renormalization 
of the effective operator that defines $J^\prime$ captures enhancements 
that are due to the near collinear fragmentation of the outgoing jet. 
These contributions in the full theory (i.e. before factorization) 
are generated by genuine collinear {\it infrared} singularities. 
Although these singularities cancel upon adding real and virtual 
contributions in a full calculation, their finite terms remain and these 
are captured by the factorization theorem. So, collinear enhancements 
in the full theory are associated with ultraviolet singlularities 
in the effective jet operator and its ensuing anomalous dimension.
In addition to these singularities, though, 
there are genuine {\it ultraviolet} singularities in the 
full theory that lead to the renormalization of its parameters, i.e. 
the coupling constant. Such ultraviolet singularies can also be present 
in the effective operator and these are not related to collinear 
enhancements but to the renormalization of the coupling constant 
in the jet function.
If one is not careful to subtract these genuine ultraviolet 
singularities from the renormalization of the operator then one is 
in danger of double counting them when the renormalization group analysis 
is performed. In other words, the UV singularities of $J^\prime$ generate 
both the collinear enhancements and the renormalization of the effective 
coupling of the jet. This effect is not seen in low orders in $F_2$ 
because the jet function $J$ starts at ${\cal O}(\alpha_s^0)$, whereas 
in $F_2$ it starts at ${\cal O}(\alpha_s)$.
We shall return to this issue when we construct explicitly the anomalous 
dimension of $J^\prime$. 

\subsection{The RG equation for the new jet function and exponentiation} 

The renormalization group equation for the jet and soft functions as well as 
the exponentiation of the logarithmic enhancements near the elastic limit 
are both consequences of the factorization formula. Therefore, we rewrite it 
here by making explicit the arguments that the participating functions 
depend on. Call $\bar{p}$ the total momentum of the $\bar{v}$-jet and 
$(1-z)$ the longitudinal momentum fraction in the $v$ direction that the jet
carries into the final state.  From the kinematics, $x \le z \le 1$,
and $\bar{p}^2 =(1-z) Q^2/x$ is the small invariant mass of the jet.  
The jet has been defined with respect to the $\bar{v}$ direction 
hence the large scale on which it depends is $(\bar{p} \cdot v)^2$. 
Then the $F_L$ factorization formula reads
\begin{eqnarray}
F_L(x, Q^2, \epsilon) &=&  
\left| H_L\left(\frac{(p\cdot \bar{v})^2}{\mu^2},
\frac{(\bar{p}\cdot v)^2}{\mu^2},\alpha_s(\mu^2) \right) \right| ^2 \, 
\int_x^1 \frac{d z}{z}  \, 
J^{\prime}\left(\frac{(1-z) Q^2}{x \mu^2}, \frac{(\bar{p} \cdot v)^2}{\mu^2},
 \alpha_s(\mu^2) \right) 
\nonumber \\
&\ & \times 
\left( V \otimes \phi \right) 
\left(\frac{z Q^2}{x \mu^2}, \frac{((p_f-\bar{p})\cdot v)^2}{\mu^2}, 
\frac{(p \cdot \bar{v})^2}{\mu^2}, \alpha_s(\mu^2), \epsilon\right).  
\label{flfacexplicit} 
\end{eqnarray}
Recall that $p_f$ is the total momentum of the final state and that  
all functions are renormalized at scale $\mu^2$. The only singularities 
present are the collinear ones with respect to the initial $v$ direction. 
In moment space the above convolution becomes a simple product:
\begin{eqnarray}
\tilde{F}_L(N, Q^2, \epsilon) &=& 
\left|H_L\left(\frac{(p\cdot \bar{v})^2}{\mu^2},
\frac{(\bar{p}\cdot v)^2}{\mu^2},\alpha_s(\mu^2) \right)\right|^2 
\nonumber \\ 
&\ & \hspace{-2.5cm} \times \frac{1}{N} 
\tilde{J}^\prime \left( \frac{Q^2}{N \mu^2}, 
\frac{(\bar{p} \cdot v)^2}{\mu^2}, \alpha_s(\mu^2) \right) \, 
\tilde{V}\left( \frac{Q^2}{N^2 \mu^2},  \alpha_s(\mu^2)\right) \, 
\tilde{\phi}\left( \frac{Q^2}{N^2 \mu^2}, \frac{(p \cdot \bar{v})^2}{\mu^2}, 
\alpha_s(\mu^2), \epsilon \right),  
\label{flfacconv}
\end{eqnarray} 
where the Mellin transformation has been defined as usual 
\begin{eqnarray}
\tilde{F_L}(N, Q^2, \epsilon) &=& \int_0^1 dx \, x^{N-1} \, 
F_L(x, Q^2, \epsilon), 
\nonumber \\  
\tilde{J}^\prime\left( \frac{Q^2}{N \mu^2}, \frac{(\bar{p} \cdot
v)^2}{\mu^2}, 
\alpha_s(\mu^2)\right) 
&=& \frac{1}{N} \int_0^1 dz \, z^{N-1} \, 
J^\prime \left(\frac{(1-z) Q^2}{x \mu^2}, \frac{(\bar{p} \cdot v)^2}{\mu^2}, 
 \alpha_s(\mu^2) \right), 
\label{Mellindef}
\end{eqnarray}
and similarly for $\tilde{V}$ and $\tilde{\phi}$. 
Note that we have not included the overall $1/N$ factor into the definition 
of $\tilde{J}^\prime$. This function will contain only the logarithmic 
in $N$ terms. Also note that the UV dominated hard scattering function 
$|H_L|^2$ contains no dependence on $N$. Its presence is necessary, however, 
to render the right-hand side of Eq.~(\ref{flfacconv}) independent of $\mu^2$.
    
Following the approach of Collins, Soper and Sterman, 
(for review and references see ref.~\cite{CoLaSte})
we can write down the two equations that the jet function satisfies. 
The first is the covariance equation
\begin{equation} 
\frac{\partial}{\partial \ln (\bar{p}\cdot v)^2} 
\ln \tilde{J}^\prime \left(\frac{Q^2}{N \mu^2}, 
\frac{(\bar{p}\cdot v)^2}{\mu^2},\alpha_s(\mu^2) \right) = 
K\left(\frac{Q^2}{N \mu^2}, \alpha_s(\mu^2)\right) 
+ G\left( \frac{(\bar{p} \cdot v)^2}{\mu^2}, \alpha_s(\mu^2) \right). 
\label{cov}
\end{equation} 
Since $F_L$ is invariant upon variation of the light cone directions with 
respect to which the jet has been defined, the jet variations $K$ and $G$ 
are cancelled by the corresponding variations of the soft function
$\tilde{V}$ 
and the hard function $|H_L|^2$ respectively. 
The dependence of $\tilde{V}$ on the light cone directions has been 
left implicit in Eq.~(\ref{flfacconv}). The quark distribution function 
$\tilde{\phi}$ is by definition independent of the final state direction 
$\bar{v}$ so it does not play a r\^{o}le in Eq.~(\ref{cov}).   
The function $K$ is defined at scale $\mu^2$ and depends only on the common 
arguments of $\tilde{J}^\prime$ and $\tilde{V}$.
The sum $K+G$ is renormalization group invariant.  
The renormalization group equation for $K$ reads 
\begin{equation}
\frac{d}{d \ln \mu^2} 
K\left( \frac{Q^2}{N \mu^2}, \alpha_s(\mu^2) \right) = 
-\frac{1}{2}\gamma_K(\alpha_s(\mu^2)), 
\label{rgK}
\end{equation}
with the well known cusp anomalous dimension in $\overline{\rm MS}$ scheme 
\cite{KodTrent}
\begin{equation}
\gamma_K(\alpha_s) = \frac{\alpha_s}{\pi} C_F 
+\left(\frac{\alpha_s}{\pi}\right)^2\left[ C_F C_A 
\left(\frac{67}{36}-\frac{\pi^2}{12}\right) -\frac{5}{18} C_F N_f\right]
+{\cal O}(\alpha_s^3).  
\label{cusp}
\end{equation}

The second equation that the jet function satisfies is its own
renormalization 
group equation 
\begin{equation} 
\frac{d}{d \ln \mu^2} 
\ln \tilde{J}^\prime \left(\frac{Q^2}{N \mu^2}, 
\frac{(\bar{p}\cdot v)^2}{\mu^2}, \alpha_s(\mu^2) \right) 
= -\frac{1}{2}\gamma_{J^\prime}(\alpha_s(\mu^2)). 
\label{rg}
\end{equation}
We employ $\overline{\rm MS}$ scheme which is a massless scheme, so the 
anomalous dimensions $\gamma_K$ and $\gamma_{J^\prime}$ can depend only on 
the renormalized coupling.

The solution of the covariance equation (\ref{cov}) is 
\begin{equation}
\ln \frac{ \tilde{J}^\prime \left(\frac{Q^2}{N \mu^2}, 
\frac{(\bar{p}\cdot v)^2}{\mu^2}, \alpha_s(\mu^2) \right) }
{ \tilde{J}^\prime \left(\frac{Q^2}{N \mu^2}, 
\frac{Q^2}{N \mu^2}, \alpha_s(\mu^2) \right)}  
=
-\frac{1}{2} \int_{Q^2/N}^{(\bar{p}\cdot v)^2} 
\frac{d \bar{\mu}^2}{\bar{\mu^2}} \left[
\ln \frac{(\bar{p}\cdot v)^2}{\bar{\mu}^2} A(\alpha_s(\bar{\mu}^2))  
+B(\alpha_s(\bar{\mu}^2)) \right],
\label{newcovsol}
\end{equation}
where we have used Eq.~(\ref{rgK}) and have defined
\begin{eqnarray}
A(\alpha_s) &=& 
\gamma_K(\alpha_s) +2 \beta(\alpha_s) 
\frac{\partial}{\partial \alpha_s} K(1, \alpha_s),
\\
B(\alpha_s)/2 &=& K(1, \alpha_s) + G(1, \alpha_s).
\label{ABdef}
\end{eqnarray}
with
\begin{equation}
\beta(\alpha_s(\mu^2)) = \frac{d}{d \ln \mu^2} \alpha_s(\mu^2)
= -\alpha_s \left[ \left(\frac{\alpha_s}{4 \pi}\right) \beta_1 
+ \left( \frac{\alpha_s}{4\pi}\right)^2 \beta_2 + ... \right].  
\label{betadef}
\end{equation} 
The solution of both the covariance and the RG equation for the jet
$J^\prime$ 
is
\begin{eqnarray}
\ln \frac{ \tilde{J}^\prime \left(\frac{Q^2}{N \mu^2}, 
\frac{(\bar{p}\cdot v)^2}{\mu^2}, \alpha_s(\mu^2) \right) }
{ \tilde{J}^\prime \left(1, 1, \alpha_s((\bar{p}\cdot v)^2) \right)} 
&=& 
-\frac{1}{2} \int_{Q^2/N}^{(\bar{p}\cdot v)^2} 
\frac{d \bar{\mu}^2}{\bar{\mu^2}} 
\left[
\ln \frac{(\bar{p}\cdot v)^2}{\bar{\mu}^2} 
\, A(\alpha_s(\bar{\mu}^2)) + B(\alpha_s(\bar{\mu}^2)) \right]
\nonumber \\ 
&\quad& 
-\int_{Q^2/N}^{(\bar{p}\cdot v)^2}  \frac{d \bar{\mu}^2}{\bar{\mu^2}}  
\, \beta(\alpha_s(\bar{\mu}^2)) \frac{\partial}{\partial \alpha_s} 
\ln \tilde{J}^\prime(1,1,\alpha_s(\bar{\mu}^2))
\nonumber \\
&\quad&
-\frac{1}{2} \int_{Q^2/N}^{\mu^2} \frac{d \bar{\mu}^2}{\bar{\mu^2}} \, 
\gamma_{J^\prime}(\alpha_s(\bar{\mu}^2)),
\label{newcovsolrg}
\end{eqnarray}
and the boundary condition for the jet function is 
(setting $(\bar{p}\cdot v)^2=Q^2$)  
\begin{equation}
\tilde{J}^\prime(1,1,\alpha_s(Q^2)) = C_F \frac{\alpha_s(Q^2)}{\pi}
+ {\cal O}(\alpha_s^2(Q^2)).
\label{initJprime}
\end{equation}

The renormalization group invariance of $F_L$ imposes the condition
\begin{equation}
\gamma_H +\gamma_{J^\prime} + \gamma_{V} +\gamma_{\phi} =0, 
\label{rgconstr}
\end{equation} 
where the anomalous dimensions of all the functions involved are defined 
just as in Eq.~(\ref{rg}). The above relation allows us to write 
$\tilde{F}_L$, Eq.~(\ref{flfacconv}), in a variety of exponentiated forms 
depending on which anomalous dimensions we wish to include in the exponent 
and what boundary conditions we choose for the factors 
$H_L$, $\tilde{J}^\prime$, $\tilde{V}$ and $\tilde{\phi}$. 
The form that is suitable for our purposes is the one that involves 
the jet anomalous dimension $\gamma_{J^\prime}$. Explicitly,
\begin{eqnarray}
 \tilde{F}_L(N, Q^2, \epsilon) & = & 
\frac{1}{N} \, |H_L(1,1,\alpha_s(Q^2))|^2 \, 
\tilde{J}^\prime(1,1,\alpha_s(Q^2)) \, 
(\tilde{V} \cdot \tilde{\phi})(1/N, \alpha_s(Q^2), \epsilon) 
\nonumber \\
&\times& 
\exp \left[-\frac{1}{2} \int_{Q^2/N}^{Q^2} 
\frac{d \mu^2}{\mu^2} 
\left( \ln \frac{Q^2}{\mu^2} A(\alpha_s(\mu^2)) 
 + B(\alpha_s(\mu^2)) \right) \right]
\nonumber \\
&\times& 
\exp \left[-\frac{1}{2} \int_{Q^2/N}^{Q^2} 
\frac{d \mu^2}{\mu^2} \left( 
\gamma_{J^\prime}(\alpha_s(\mu^2))      
+2 \beta(\alpha_s(\mu^2)) \frac{\partial}{\partial \alpha_s} 
\ln \tilde{J}^\prime(1,1,\alpha_s(\mu^2)) 
\right) \right] 
\nonumber \\
&+& {\cal O}\left(\frac{1}{N} \ln^0N \right). 
\label{flsol}
\end{eqnarray}
We shall be referring to this equation as the Sudakov exponentiated form of 
$\tilde{F}_L$. 
By inspecting the above expression it is easy to see where the logarithmic 
in $N$ corrections come from and how they are organized. As usual, the
leading 
$\ln N$ terms arise from the soft regions and are contained in the function 
$A$. The logarithmic corrections due to the fragmentation of the final state 
jet are contained in $\gamma_{J^\prime}$ and the ones due to the 
renormalization of the coupling constant in the jet function are in beta 
function term. The beta function term arises from the evolution of the 
boundary condition from the low scale $\tilde{J}^\prime(1,1,\alpha_s(Q^2/N))$ 
to the high scale  $\tilde{J}^\prime(1,1,\alpha_s(Q^2))$. 
It is worth empasizing again here that in $F_L$ even the lowest order 
contribution cannot be normalized unambiguously because it starts 
at ${\cal O}(\alpha_s)$. 
Since the renormalizability of the theory determines the structure of the 
beta function term in the exponent, we can absorb it into the 
anomalous dimension $\gamma_{J^\prime}$ as 
\begin{equation}
\gamma^\prime_{J^\prime}(\alpha_s) = \gamma_{J^\prime}(\alpha_s) 
+2  \beta(\alpha_s) \frac{\partial}{\partial \alpha_s} 
\ln \tilde{J}^\prime(1,1,\alpha_s), 
\label{gammaprime}
\end{equation} 
while keeping the boundary condition at the high scale, i.e. 
$\tilde{J}^\prime(1,1,\alpha_s(Q^2))$. The practical advantage of this 
simple redefinition is that when we compute the anomalous dimension of the 
operator in Eq.~(\ref{Jprimedef}) we do not have to subtract out 
the UV poles that renormalize the coupling as was suggested at the end 
of the previous section. In this fashion we compute $\gamma^\prime_{J^\prime}$
and not $\gamma_{J^\prime}$. Once the distinction between the two is made 
as in Eq.~(\ref{gammaprime}), there is no danger of overcounting.  
This is the approach that we follow in the next section. 
  
Not all logarithmic in $N$ dependence is contained in the 
exponential of Eq.~(\ref{flsol}), but it resides also in the 
$\tilde{V} \cdot \tilde{\phi}$ factor. This can be captured too, 
if the corresponding anomalous dimensions are introduced in the exponent. 
We shall leave this factor as is, because it is common to $\tilde{F_L}$ and 
$\tilde{F_2}$ and contains no further information that is characteristic of 
the longitudinal structure function only.  
Our final objective is to connect the exponentiated form of $\tilde{F}_L$,
Eq.~(\ref{flsol}), with the operator product expansion factorization 
as given by Eq.~(\ref{OPEfac}). 
In the $\overline{\rm MS}$ scheme the coefficient 
function in moment space and at scale $\mu^2=Q^2$ is given by 
\begin{equation}
\tilde{C}_L(N, Q^2/\mu^2=1) = \frac{\tilde{F}_L(N, Q^2, \epsilon)}
{\tilde{f}(N, Q^2, \epsilon)}.  
\label{barC}
\end{equation} 
The regularized quark distribution function $\tilde{f}(N, Q^2, \epsilon)$ 
can also be written as an exponential by simply integrating its DGLAP 
evolution equation in $Q^2$, given one boundary condition. 
This is done in the presence of the regulator, 
i.e. in $D=4-2 \epsilon$ dimensions. 
It is useful to choose for boundary condition the normalization
\cite{CoLaSte} 
\begin{equation}
\tilde{f}(N, Q^2=0, \epsilon) = 1, 
\label{fnorm}
\end{equation}  
which yields the solution 
\begin{equation}
\tilde{f}(N, Q^2, \epsilon) = 
\exp \left[ -\int_0^{Q^2} \frac{d \mu^2}{\mu^2} 
\Gamma_{qq}(N, \epsilon, \alpha_s(\mu^2)) \right]. 
\label{fsol}
\end{equation}
Here $\Gamma_{qq}$ is the Mellin transformation of the (singular) quark 
splitting function $P_{qq}(x, \epsilon, \alpha_s(\mu^2))$. 
It is important to keep in mind that the expression in Eq.~(\ref{fsol}) 
is defined only in the presence of the regulator $\epsilon$. 
The integration over the coupling constant down to $\mu^2=0$ is also 
defined by using a dimensionally continued $\alpha_s$ \cite{CoLaSte}. 

When we attempt to compute the coefficient function $\bar{C}_L$ from   
the ratio in Eq.~(\ref{barC}) we encounter the following two problems. 
First, we cannot use the exponentiated form of Eq.~(\ref{flsol}) because 
Sudakov factorization organizes the exponent of $\tilde{F}_L$ in terms of 
logarithmic in $N$ dependence whereas the DGLAP evolution organizes the 
exponent of $\tilde{f}$ in terms of logarithms of $Q^2$. 
In other words the ratio $\tilde{F}_L/\tilde{f}$ is not a simple ratio of two 
exponentials because parts of the prefactors in the Sudakov exponentiation 
of $\tilde{F}_L$ are contained in the DGLAP exponent of $\tilde{f}$. 
The consistent way to compute the coefficient function is to write also 
$\tilde{F}_L$ as a single exponential by solving its own DGLAP equation 
in $D= 4-2\epsilon$ dimensions. This is indeed what is done for
$\tilde{F}_2$. 
The boundary condition used in the $\tilde{F}_2$ DGLAP equation is the 
normalization
\begin{equation}
\tilde{F}_2(N, Q^2=0, \epsilon) = 1, 
\label{F2norm}
\end{equation} 
and the solution is again a single exponential, like in Eq.~(\ref{fsol}). 
Then the coefficient function $\tilde{C}_2$ follows simply from 
Eq.~(\ref{barC}) with $L \rightarrow 2$.
It is here that we encounter the second problem for the longitudinal 
strucutre function. $\tilde{F}_L$ cannot be normalized as $F_2$ above, 
because it starts at ${\cal O}(\alpha_s)$~\footnote{If we were to use the 
analytically continued $\alpha_s$ that 
would give $\tilde{F}_L(N, Q^2=0, \epsilon)=0$.}.

Although we cannot treat formally $\tilde{F}_L$ like $\tilde{F}_2$, we 
observe that a large part in their respective Sudakov exponentiated forms is 
the same.  
Indeed, if we repeat the previous renormalization group analysis for 
$\tilde{F}_2$, we get an answer for $\tilde{F}_2$ which 
can be obtained from Eq.~(\ref{flsol}) by simply substituting 
$F_L, \, H_L \rightarrow F_2, \,  H_2$  and $J \rightarrow J^\prime$.  
$\tilde{F}_2$ can be normalized as in Eq.~(\ref{F2norm}) because 
$J=1+{\cal O}(\alpha_s)$. 
Therefore, if we know the coefficient function $\bar{C}_2$ to some order 
in $\alpha_s$ we can predict the coefficient function $\bar{C}_L$ to one
order higher by using the Sudakov factorization. 
All we need to do is to isolate the parts that are different between 
$F_2$ and $F_L$ in their exponentiated forms.
The problem of normalization of $\tilde{F}_L$ at $Q^2=0$ is solved by 
defining a function $\tilde{\cal F}_L$ as 
\begin{equation}
\tilde{\cal F}_L(N, Q^2, \epsilon) = N 
\frac{\tilde{F}_L(N, Q^2, \epsilon)}{J^\prime(1,1, \alpha_s(Q^2))}. 
\label{scriptf}
\end{equation} 
Since the jet has been moded out,  we can normalize the ratio as 
$\tilde{\cal F}_L(N, Q^2=0, \epsilon) =1$. 
Then we compare the Sudakov exponentiation of $\tilde{F}_L$ with the 
corresponding one for $\tilde{F_2}$ and we identify the parts in the 
exponent that are different. These parts are the corrections that 
must be included in $\bar{C}_2$ to obtain $\bar{C}_L$. 
Once we do this we restore $\tilde{F}_L$ via Eq.~(\ref{scriptf}). 
The net result of this manipulation is that the 
coefficient function for $\tilde{C}_L$ is predicted to be 
\begin{equation}
\tilde{C}_L = \frac{\tilde{J}^\prime(1,1,\alpha_s(Q^2))}{N} 
\frac{\tilde{\cal F}_L}{\tilde{F}_2} \tilde{C}_2, 
\label{allorderC}
\end{equation}
and the ratio of the two structure functions can be readily obtained 
from their Sudakov exponentiated forms. 
The net effect of the ratio $\tilde{\cal F}_L/\tilde{F}_2$ is to remove 
from $\tilde{C}_2$ the contributions coming from the jet $J$ of $F_2$ 
and replace them with the contributions from the $J^\prime$ jet of $F_L$.   
To ${\cal O}(\alpha_s^2)$ the above relation yields 
\begin{equation}
\tilde{C}^{(2)}_L(N, Q^2/\mu^2= 1) = 
\frac{1}{N} C_F 
\left(\gamma_K^{(1)} \ln^2 N +\frac{1}{2}\gamma^{\prime(1)}_{J^\prime} 
\ln N \right) 
+{\cal O}\left( \frac{1}{N} \ln^0 N \right).
\label{barCpredict}
\end{equation}

\section{Explicit calculations to ${\cal O}(\alpha_s^2)$ }
\setcounter{equation}{0}

In this section we prove via explicit calculations the $F_L$ 
factorization formula Eq.~(\ref{flfac}) to  ${\cal O}(\alpha_s^2)$,
or equivalently that Eq.~(\ref{barCpredict}) holds. 
This is done by comparing our results with 
the coefficient functions $\bar{C}_L^{(n)}(x,1)$, $n=1,2$ in 
Eqs.~(\ref{Cl1}, \ref{cl2}). 
The new object to be computed is the jet function $J^\prime$ 
defined in Eq.~(\ref{Jprimedef}), since 
both the soft function $V$ and the quark distribution $\phi$ are the same 
as in $F_2$ and can be taken from the literature. We show that the singular 
$\ln(1-x)$ fragmentation terms in $\bar{C}_L^{(2)}(x,1)$  do arise from 
the UV renormalization of the operator defining $J^\prime$ and we construct 
its anomalous dimension $\gamma^\prime_{J^\prime}$ to ${\cal O}(\alpha_s)$. 

The analysis of subsection 3.2 resulted in a set of Feynman rules for 
computing the logarithmic enhancements of the new jet function $J^\prime$. 
Specifically, the coupling of gluons from the $\bar{v}$-jet to the 
external lines is given by the effective non-local vertex $O^\lambda(l)$ 
defined in Eq.~(\ref{Odef}). In that expression, though, the transverse and 
light cone polarizations have been separated, whereas we find it 
more convenient for our calculations to compute traces with covariant 
objects. To this end we define the covariant vertex $O^\lambda(l)_{\rm cov}$
as
\begin{equation}
O^\lambda(l)= O^\lambda(l)_{\rm cov}
-\left(\bar{v}^\lambda-\frac{\bar{v}\cdot l}{v\cdot
l}v^\lambda\right)\not{v}, 
\ \ \
O^\lambda(l)_{\rm cov} =
\left(\gamma^\lambda-\frac{v^\lambda}{v\cdot l}\not{l} \right).
\label{Ocov}
\end{equation}
Then
\begin{equation}
{\rm tr}\left[\not{p} O^\lambda(l) ...\right] 
={\rm tr} \left[ \not{p} O^\lambda(l)_{\rm cov}...\right]
\label{Otr}
\end{equation}
so we can use the covariant vertex for computing traces.  
Note that in terms of the operator definitions, $O_{\rm cov}$ is generated 
by the operator in the numerator of Eq.~(\ref{Jprimedef}) after performing 
a gauge transformation in the light cone direction. 
All functions entering the factorization formula have been defined in a gauge 
independent way and from now on we employ the Feynman gauge. 
Finally, in the $\overline{\rm MS}$ scheme it is convenient for presenting 
intermediate results to define the ratio of scales 
\begin{equation} 
S=\frac{e^{-\gamma_E} 4 \pi x \mu^2}{Q^2}. 
\label{Sdef}
\end{equation}

\subsection{The ${\cal O}(\alpha_s)$ contribution to $J^{\prime}$} 

To first order in the coupling $F_L$ has no logarithmic singularities, 
the answer being simply 
\begin{equation}
F_L^{(1)}(x, Q^2, \epsilon) = c_L^{(1)} =  C_F x.  
\label{fllowest}
\end{equation}
The $F_L$ factorization formula to this order reads 
\begin{equation}
F_L^{(1)}(x, Q^2, \epsilon) = J^{\prime (1)}(x, Q^2) \otimes F_2^{(0)} 
= J^{\prime(1)}(x, Q^2) \, . 
\label{flfac1}
\end{equation}  
We check that $J^{\prime(1)}$ obtained from 
Eq.~(\ref{Jprimedef}) comes with the correct normalization to reproduce the 
above lowest order result. There is only one diagram to compute, that of 
fig.~4. From Eqs.~(\ref{wtens}) and (\ref{PS2fin}) we have 
\begin{eqnarray}
J^{\prime(1)}(x,Q^2) &=& 
\left(\frac{2 x^2}{\alpha_s Q^2}\right) (4\pi\alpha_s) 
\mu^{2\epsilon} C_F \frac{1}{2} \int d PS_2 \, 
{\rm tr}\left[ O^\lambda(k_2) \not{k}_1 O^\rho(k_2) \not{p} \right]
\left( -\eta_{\lambda \rho} \right)
\nonumber \\ 
&=& C_F x \left(\frac{1-\epsilon}{1-2\epsilon}\right)
\frac{\Gamma(1-\epsilon)}{\Gamma(1-2\epsilon)} 
\left(\frac{4\pi x\mu^2}{(1-x)Q^2}\right)^\epsilon 
\nonumber \\ 
&=& C_F x S^\epsilon  
[1-2\epsilon(\ln(1-x)-1)]
= F_L^{(1)}(x, Q^2,\epsilon)
\label{fl1}
\end{eqnarray}
We note that the whole contribution to $F_L^{(1)}$ comes from $J^{\prime(1)}$ 
and no IR subtractions are needed to this order. For convenience 
we have included in $J^{\prime(1)}$ all overall normalizing factors 
associated with $F_L$. 
This short calculation shows that $J^{\prime}$ as 
defined by Eq.~(\ref{Jprimedef}) is correctly normalized. 

\subsection{The ${\cal O}(\alpha_s^2)$ contribution to the jet function 
$J^\prime$} 

The non-trivial infrared behavior of $F_L$ starts at ${\cal O}(\alpha_s^2)$. 
We compute the jet function to this order, $J^{\prime(2)}$, and show that 
after summation over final state cuts and IR subtractions as required by 
the denominator of Eq.~(\ref{Jprimedef}) the only surviving singularities 
for the operator diagrams are simple UV poles. 
In section 3.3 we argued that the $J^\prime$ operator can be multiplicatively 
renormalized in moment space as 
\begin{equation}
\tilde{J}^\prime_u(N,Q^2,\alpha_s(\mu^2),\epsilon) = 
Z_{J^\prime}(\alpha_s(\mu^2), \epsilon) \, 
\tilde{J}^\prime(N, Q^2,\alpha_s(\mu^2)). 
\label{Jprimeren}
\end{equation}
In a massless scheme such as $\overline{\rm MS}$, the anomalous dimension 
$\gamma^\prime_{J^\prime}$ can be constructed from the renormalization 
constant 
$Z_{J^\prime}$ as 
\begin{equation}
\gamma^\prime_{J^\prime}(\alpha_s(\mu^2)) = - 2 \alpha_s 
\frac{\partial}{\partial \alpha_s} 
{\rm Res} Z_{J^\prime}(\alpha_s(\mu^2), \epsilon). 
\label{gammams}
\end{equation}
Equivalently, to the order we compute, and because of the relations 
\begin{eqnarray}
\int_0^1 dx \, x^{N-1} \ln(1-x) &=& -\frac{\ln N}{N}+{\cal O}(\ln^0N/N),
\\ 
\int_0^1 dx \, x^{N-1} \ln^2(1-x) &=& 
\frac{1}{N} (\ln N+\gamma_E)^2 +{\cal O}(\ln^0N/N), 
\label{momlog}
\end{eqnarray} 
we can directly generate the 
${\cal O}(\ln(1-x))$ terms for $J^\prime$ in $x$-space. Their coefficient 
is the  sign-inverted residue of the UV singular part of $J^{\prime(2)}$. 

There are seven  diagrams that determine $J^{\prime(2)}$ and they are 
depicted in fig.~5.
Since certain subtleties occur in the calculation arising 
from the presence of both linear (eikonal) and quadratic 
denominators in the loop integrations we proceed by analyzing each diagram 
in turn.
In the full calculation to this order there are fourteen diagrams for $F_L$.  
Whenever a diagram in our effective approach is free from 
IR divergences we provide direct comparison with the corresponding one 
of the full calculation taken from ref. \cite{DDKS}.
At the practical level, the effective approach is vastly more economical 
in capturing the logarithmic enhancements. 

\medskip

\noindent
{\it Diagram $(a)$}: 
\begin{equation}
(a) = 2 i(4 \pi \alpha_s)^2 \,C_F^2 \,\frac{1}{2}\,\mu^{4\epsilon} 
\int dPS_2 \,
\int \frac{d^D l}{(2 \pi)^D} \, \frac{N_{(a)}}{(l-\bar{p})^2 \, l^2 },  
\label{aini}
\end{equation} 
where a factor of 2 has been included from left$\leftrightarrow$right equal
contributions. The numerator factor is 
\begin{eqnarray} 
N_{(a)} &=& \frac{1}{g_s^2}{\rm tr}\left[O^\rho(k_2) \not{k_1} \gamma_\rho 
\frac{1}{\not{\bar{p}}} \gamma_\lambda (\not{\bar{p}}-\not{l}) O^\lambda(l) 
\not{p}\right],
\nonumber \\ 
&=& 2 \left(1 -  \frac{p \cdot {\bar{p}}}{p\cdot l}\right) 
\frac{1}{g_s} {\rm tr}\left[ O^\rho \not{k}_1 \gamma_\rho 
\frac{1}{\not{\bar{p}}} \not{l} \not{p}\right] 
\label{na}
\end{eqnarray}
and the loop integrals yields through the use of Eq.~(\ref{cool2}) in 
the appendix 
\begin{eqnarray}
\mu^{2\epsilon}\int \frac{d^D l}{(2 \pi)^D} \, 
\frac{N_{(a)}}{(l-\bar{p})^2 \, l^2 } 
&=& \frac{1}{g_s}
{\rm tr}\left[ O_{\rm cov}^\rho \not{k}_1 \gamma_\rho\not{p}\right] 
\frac{-i}{(4 \pi)^2}\frac{1}{\epsilon} 
=-2\cdot 4 \, (p\cdot k_1) \frac{-i}{(4 \pi)^2}\frac{1}{\epsilon}
+{\cal O}(\epsilon^0).  
\label{loopa}
\end{eqnarray}
Then the diagram becomes 
\begin{equation}
(a) = -\alpha_s^2 \, C_F^2 \, \frac{1}{\epsilon} \frac{Q^2}{2 \pi x} 
\int_0^1 d \frac{p \cdot k_1}{p \cdot \bar{q}} 
\frac{p \cdot k_1}{p \cdot \bar{q}} +{\cal O}(\epsilon^0) \
= \  -\frac{\alpha_s^2 Q^2}{4 \pi x} C_F^2 \frac{1}{\epsilon} 
+{\cal O}(\epsilon^0).  
\label{afin}
\end{equation}
By including the overall factor from Eq.~(\ref{wtens}) we obtain the
contribution to the unrenormalized $J^{\prime(2)}_u$ 
\begin{equation}
\left(\frac{\alpha_s}{\pi}\right)^2 J^{\prime (2)}_{u(a)} = 
\frac{1}{4 \pi} \frac{8 x^2}{Q^2} (a) = 
-\left(\frac{\alpha_s}{ \pi}\right)^2 \frac{1}{2} C_F^2 x \frac{1}{\epsilon}
+{\cal O}(\epsilon^0) 
\label{J2aren}
\end{equation}
This leads to the anomalous dimension 
\begin{equation}
\gamma_{J^\prime(a)}(\alpha_s) =  \frac{\alpha_s}{\pi} C_F, 
\label{gammaJ2a}
\end{equation}
as well as to the contribution for the renormalized jet function 
\begin{equation}
J^{\prime(2)}_{(a)} =  \frac{1}{2} C_F^2 \ln(1-x) + {\rm reg.}
\label{J2a}
\end{equation}  

\medskip

\noindent 
{\it Diagram $(b)$}: 
\begin{equation}
(b) =- 2(4 \pi \alpha_s)^2 \, \left(\frac{iC_F C_A}{2}\right) \, 
\frac{1}{2} \, \mu^{4\epsilon} \int dPS_2 \, \int \frac{d^D l}{(2 \pi)^D} \, 
\frac{N_{(b)}}{(\bar{p}-l)^2\,(l-k_2)^2\,l^2}, 
\label{bini}
\end{equation} 
and the numerator factor is 
\begin{eqnarray}
N_{(b)} &=& \frac{1}{g_s^2}{\rm tr} \left[ O^\rho(k_2) \not{k}_1
\gamma^\sigma 
(\not{\bar{p}}-\not{l}) O^\lambda(l) \not{p} \right] 
V_{\rho \sigma \lambda}(-k_2, k_2-l, l), 
\nonumber \\
&\rightarrow & 
16 (p \cdot k_1) l^2 \left(1-\frac{p \cdot k_1}{p \cdot l} \right).  
\label{nbfin}
\end{eqnarray} 
In the second step  only terms  that lead to UV singularities have been 
retained. There is a subtlety involved in the derivation of the above 
expression. A term proportional to $l_\sigma l_\rho$ is to be replaced as 
 $l_\sigma l_\rho \rightarrow (1/2) l^2 \eta_{\sigma \rho}$. Note the 
factor 1/2 instead of the usual 1/4. The latter case is valid  for the UV part
of a loop integral {\it if} the integrand contains only quadratic 
denominators. In our case, however,  it also contains a linear (eikonal) 
denominator and one must be careful not to overcount the polarizations that 
contribute to the UV divergent part (2 and not 4). 
Then, through the use of Eq.~(\ref{cool1}) in the appendix we obtain
\begin{eqnarray}
(b) &=& -2(4 \pi \alpha_s)^2 \alpha_s^2 \, \left(\frac{iC_F C_A}{2}\right) \, 
\frac{1}{2}\, \mu^{2\epsilon} \int dPS_2 \,16(p\cdot k_1)
\frac{i}{(4 \pi)^2} \left(1+\ln\frac{p\cdot k_2}{p\cdot \bar{q}}\right)
\frac{1}{\epsilon} +{\cal O}(\epsilon^0)
\nonumber \\
&=& -\frac{\alpha_s^2 Q^2}{4 \pi  x}  \left(\frac{C_F C_A}{2}\right) \, 
\frac{1}{\epsilon} +{\cal O}(\epsilon^0). 
\label{bfin}
\end{eqnarray} 
Similarly to diagram $(a)$, the corresponding contribution to 
the anomalous dimension is 
\begin{equation}
\gamma^\prime_{J^\prime(b)}(\alpha_s) = \frac{\alpha_s}{\pi} \frac{C_A}{2}, 
\label{gammaJ2b}
\end{equation}
and to the renormalized jet function 
\begin{equation}
J^{\prime (2)}_{(b)} =  \frac{1}{4} C_F C_A \ln(1-x)+{\rm reg.} 
\label{J2b}
\end{equation}  
Eq.~(\ref{J2b}) reproduces the result for graph (10) in \cite{DDKS}.

\medskip

\noindent 
{\it Diagram $(c)$}: 
\begin{eqnarray}
(c) &=& 2i(4 \pi \alpha_s)^2 \, \left(C_F^2-\frac{C_F C_A}{2}\right) \, 
\frac{1}{2}\,\mu^{4\epsilon} \int dPS_2 \, \int \frac{d^D l}{(2 \pi)^D} \, 
\frac{N_{(c)}}{(l-\bar{p})^2\,(l-k_1)^2\,l^2},
\label{cini}
\end{eqnarray} 
and the numerator factor can be approximated in the UV as  
\begin{eqnarray}
N_{(c)} &=& \frac{1}{g_s^2}
{\rm tr} \left[ O^\rho(k_2) \not{k}_1 \gamma^\lambda 
(\not{k}_1-\not{l}) \gamma_\rho (\not{\bar{p}}-\not{l}) O^\lambda(l) \not{p} 
\right] 
\nonumber \\ 
&\rightarrow &  8 (p\cdot k_1) 
\left( 1-2\frac{p\cdot k_1}{p\cdot l}\right)l^2.
\label{ncfin}
\end{eqnarray} 
The same remark about counting the polarizations that contribute to the UV 
singularities as in diagram $(b)$ applies here. 
Then, through the use of Eq.~(\ref{cool1}) of the appendix we obtain  
\begin{eqnarray} 
(c)  &=& -8 \alpha_s^2 \, \left(C_F^2-\frac{C_F C_A}{2}\right) \, 
\mu^{2\epsilon} \int dPS_2 \,  (p\cdot k_1) 
\frac{1}{\epsilon}\left(1+2\frac{p\cdot k_1}{p\cdot k_2} 
\ln\frac{p\cdot k_1}{p\cdot \bar{q}}\right) 
\nonumber \\
&=& \frac{\alpha_s^2 Q^2}{\pi x}\, \left(C_F^2-\frac{C_F C_A}{2}\right)\, 
\left(\zeta(2)-\frac{3}{2}\right)\frac{1}{\epsilon} +{\cal O}(\epsilon^0), 
\label{cfin}
\end{eqnarray}
which contributes to the anomalous dimension as 
\begin{equation}
\gamma^\prime_{J^\prime(c)}(\alpha_s) = -\frac{\alpha_s}{\pi} 
4\left(C_F-\frac{1}{2}C_A\right)\left(\zeta(2)-\frac{3}{2}\right),
\label{gammaJ2c}
\end{equation}
and to the renormalized jet function as
\begin{equation}
J^{\prime (2)}_{(c)} = 
-2 \left(C_F^2-\frac{C_F C_A}{2}\right)\,\left(\zeta(2)-\frac{3}{2}\right)
 \ln(1-x)+{\rm reg.} 
\label{J2c}
\end{equation}  
Eq.~(\ref{J2c}) reproduces the result for graph (13) in \cite{DDKS}.

So far, the three graphs we have computed are in one-to-one correspondence
with the corresponding graphs of the full theory. They are finite in the full 
theory and become UV divergent in the effective theory. 
Now we proceed to calculate the rest of the graphs. 
These are not in one-to-one correspondence with graphs of the full theory, 
which is the default case in factorization.  
No more comparison for intermediate results will be possible, only for the 
final answer. 

\medskip

\noindent 
{\it Diagrams $(d)$ and $(e)$}: 
For these two diagrams no subtleties are involved. 
Standard manipulations yield   
\begin{eqnarray}
\gamma^\prime_{J^\prime(d)}(\alpha_s) &=& \frac{\alpha_s}{\pi} \frac{C_F}{2},
\\
J^{\prime (2)}_{(d)} &=& \frac{1}{4} C_F^2 \ln(1-x) +{\rm reg.} 
\label{J2d}
\end{eqnarray}  
and
\begin{eqnarray}
\gamma^\prime_{J^\prime(e)}(\alpha_s) &=& \frac{\alpha_s}{\pi} 
\left( -\frac{5}{6} C_A +\frac{1}{3} N_f\right), 
\\
 J^{\prime (2)}_{(e)} &=& \left(-\frac{5}{12} C_F C_A 
+\frac{1}{6} C_F N_f \right) \ln(1-x) 
+{\rm reg.} 
\label{J2e}
\end{eqnarray}  

The last two diagrams $(f)$ and $(g)$ are the most interesting to this order 
because they have both UV and IR singularities. Recall that the jet function 
$J^\prime$ is by definition free from IR singularities and its UV 
renormalization generates the fragmentation $\ln (1-x)$ terms in the 
coefficient function. We must therefore show that the left over IR 
singularities in the graphs $(f)$ and $(g)$, after summing over their two and 
three particle cuts, are precisely removed by the soft function $V$. 
The remaining UV poles will contribute to the $J^\prime$ anomalous dimension. 

\medskip

\noindent 
{\it Diagram $(f)$}: \\
The two particle cut of this diagram $(f_2)$ is 
\begin{eqnarray}
(f_2) &=&  2(4 \pi \alpha_s)^2  \, \left(\frac{i C_F C_A}{2}\right) \, 
\frac{1}{2}\,\mu^{4\epsilon} \int dPS_2 \, \int \frac{d^D l}{(2 \pi)^D} \, 
\frac{N_{(f)}}{(l-k_2)^2\,l^2\,(p\cdot l)}
\nonumber \\  
&=& \frac{\alpha_s^2 Q^2}{4\pi x}\,  \left(\frac{C_F C_A}{2}\right) 
S^\epsilon (1-x)^{-\epsilon}
\nonumber \\
&\ & \times  
\left[ \frac{1}{\epsilon}+\frac{1}{(-\epsilon)} 
-2 S^\epsilon   
\left(1+\frac{5}{2}\epsilon\right) \left(\frac{1}{\epsilon(-\epsilon)} + 
\frac{1}{(-\epsilon)^2}\right) +{\cal O}(\epsilon^0)\right]. 
\label{f2fin}
\end{eqnarray} 
We have separated UV from IR singularities by retaining the sign in front of 
the regulator. 

The three particle cut $(f_3)$ can similarly be obtained through the use of 
the three massless particle phase space given in Eq.~(\ref{PS3fin}) of the 
appendix in terms of convenient coordinates for our case. The result is 
\begin{eqnarray}
(f_3)&=& 2(-i)(4\pi\alpha_s)^2 \left(\frac{iC_F C_A}{2}\right) \frac{1}{2}\, 
\mu^{4\epsilon} \int dPS_3 (2\cdot 4)(1-\epsilon) (p\cdot k_1)  
\frac{p\cdot(k_3+2k_2)}{(p\cdot k_3) \,(k_2+k_3)^2} 
\nonumber \\ 
&=&  \frac{\alpha_s^2 Q^2}{4\pi x} \left(\frac{C_FC_A}{2}\right) 
S^{2\epsilon} (1-x)^{-2\epsilon}
\left(1+\frac{9}{2}\epsilon\right) 
\left[ \frac{3}{(-\epsilon)}+\frac{2}{(-\epsilon)^2}+{\cal O}(\epsilon^0)
\right].
\label{f3fin}
\end{eqnarray} 
Adding the two and three particle cuts we obtain for the full cut diagram 
\begin{eqnarray}
(f)=(f_2)+(f_3) &=& 
\frac{\alpha_s^2 Q^2}{4\pi x} \left(\frac{C_FC_A}{2}\right)
S^\epsilon (1-x)^{-\epsilon} 
\nonumber \\
&\ & \times
\left[ \frac{1}{\epsilon}
-2 S^\epsilon 
\left(1+\frac{5}{2}\epsilon\right) \frac{1}{\epsilon(-\epsilon)} 
+2\ln(1-x)\frac{1}{(-\epsilon)}+{\cal O}(\epsilon^0) \right]. 
\label{ffin}
\end{eqnarray}
We note the cancellation of the double IR poles which arise from soft gluon 
emission. Simple IR and UV poles and their combination do survive. 

\medskip

\noindent 
{\it Diagram $(g)$}: \\ 
Similar manipulations as in diagram $(f)$ yield the following  results for 
the two and three particle cuts.
\begin{eqnarray}
(g_2) &=& 2(-i)(4 \pi \alpha_s)^2 \, \left(C_F^2-\frac{C_F C_A}{2}\right) \, 
\frac{1}{2} \, \mu^{4\epsilon} \int dPS_2 \, \int \frac{d^D l}{(2 \pi)^D} \, 
\frac{N_{(g)}}{(l-k_2)^2\,l^2\,(p\cdot l)}
\nonumber \\ 
&=& \frac{\alpha_s^2 Q^2}{2\pi x}\,  \left(C_F^2-\frac{C_F C_A}{2}\right) 
S^\epsilon (1-x)^{-\epsilon} 
\nonumber \\
&\ & \times
\left[ \frac{1}{\epsilon}+\frac{1}{(-\epsilon)} 
- S^\epsilon   
\left(1+\frac{3}{2}\epsilon\right) \left(\frac{1}{\epsilon(-\epsilon)} + 
\frac{1}{(-\epsilon)^2}\right) +{\cal O}(\epsilon^0)\right]. 
\label{g2fin}
\end{eqnarray}
\begin{eqnarray}
(g_3)&=& 2(4\pi\alpha_s)^2 \left(C_F^2-\frac{C_F C_A}{2}\right) \frac{1}{2}
\, 
\mu^{4\epsilon} \int dPS_3 \,16(1-\epsilon)  
\frac{(p\cdot k_1) (p\cdot(k_1+k_3))}{(p\cdot k_3) \,(k_1+k_3)^2} 
\nonumber \\ 
&=&  \frac{\alpha_s^2 Q^2}{2\pi x} \left(C_F^2-\frac{C_FC_A}{2}\right) 
S^{2\epsilon} (1-x)^{-2\epsilon} 
\left(1+\frac{7}{2}\epsilon\right) 
\left[\frac{1}{(-\epsilon)}+\frac{1}{(-\epsilon)^2}+{\cal O}(\epsilon^0)
\right].
\label{g3fin}
\end{eqnarray}
Adding the two and three particle cuts we obtain for the full cut diagram  
\begin{eqnarray}
(g)= (g_2)+(g_3) &=& 
\frac{\alpha_s^2 Q^2}{2\pi x} \left(C_F^2-\frac{C_FC_A}{2}\right)
S^\epsilon (1-x)^{-\epsilon}
\nonumber \\
&\ & \ \times 
\left[ \frac{1}{\epsilon}
-S^\epsilon \left(1+\frac{3}{2}\epsilon\right) 
\frac{1}{\epsilon(-\epsilon)}
+\ln(1-x) \frac{1}{(-\epsilon)} +{\cal O}(\epsilon^0) \right]. 
\label{gfin}
\end{eqnarray}
As in diagram $(f)$, we note the cancellation of the IR double poles. 
UV and IR simple poles do remain.

Finally, we add the contributions from diagrams $(f)$ and $(g)$ and 
multiply the result by $(2\pi x^2/\alpha_s^2 Q^2)$, 
see Eqs.~(\ref{alphaexp}, \ref{wtens}), to obtain their contribution to 
the unrenormalized $J^{\prime(2)}$. 
It is convenient to separate the color structures, since they behave 
quite differently. The result is
\begin{eqnarray}
J^{\prime(2)}_{(f+g)u} &=& 
 C_F^2 x
S^\epsilon (1-x)^{-\epsilon} 
\left[ \frac{1}{\epsilon}
-S^\epsilon  \left(1+\frac{5}{2}\epsilon\right) \frac{1}{\epsilon(-\epsilon)}
+\ln(1-x)\frac{1}{(-\epsilon)}+{\cal O}(\epsilon^0) \right] 
\nonumber \\
&\ & + \left(\frac{C_FC_A}{2}\right) x 
S^\epsilon (1-x)^{-\epsilon} 
\left[\frac{1}{2\epsilon} +{\cal O}(\epsilon^0) \right].
\label{f+g}
\end{eqnarray}
We observe the absence of any  IR poles in the $C_FC_A$ color structure. 

\medskip
\noindent 
{\it Total contribution to $J^{\prime (2)}$}: \\ 
To ${\cal O}(\alpha_s^2)$ the jet function $J^{\prime(2)}$ 
is obtained from the sum of the diagrams  $(a)$-$(g)$ after subtracting 
the purely soft contributions to the same order. 
The IR subtraction term for $F_L$ starts at ${\cal O}(\alpha_s^2)$ and it is 
$J^{\prime(1)}\otimes V^{(1)}$. Explicitly,
\begin{equation}
J^{\prime(2)}(x) =J^{\prime(2)}_{u}(x) -  (J^{\prime(1)}\otimes V^{(1)})(x) 
\label{renjet}
\end{equation}
where the subscript $u$ denotes that the jet function is unrenormalized 
and IR unsubtracted.
The soft function $V(1-x)$ is universal and familiar. To ${\cal
O}(\alpha_s)$  
it is defined in fig. 6. In the Feynman gauge it is  
\begin{equation}
V^{(1)}(1-x) = 
 C_F S^\epsilon 
\left[-\frac{1}{(-\epsilon)\epsilon} \delta(1-x) 
+\frac{1}{(-\epsilon)} \left(\frac{1}{1-x}\right)_+ 
+ \left(\frac{\ln (1-x)}{1-x}\right)_+ +{\cal O}(\epsilon)\right]. 
\label{V1}
\end{equation} 
The argument of $V$ is the longitudinal momentum fraction that enters 
the $\bar{v}$-jet. The first term in the square brackets is of UV origin and 
comes from the virtual contribution. Note the absence of IR double poles.
We compute the convolution $J^{\prime(1)}\otimes V^{(1)}$ using the 
integral lemmata (\ref{+1}-\ref{+5}) in the appendix. The result is 
\begin{eqnarray}
(J^{\prime(1)}\otimes V^{(1)})(x) &=& 
J^{\prime(1)}(x)\otimes V^{(1)}(1-x) 
\nonumber \\
&=& C_F^2 S^{2\epsilon}\left[ x-\epsilon x(\ln(1-x)-1)\right] 
\nonumber \\
&\ & \ \ 
\otimes 
\left[-\frac{1}{(-\epsilon)\epsilon} 
\delta(1-x) +\frac{1}{(-\epsilon)} \left(\frac{1}{1-x}\right)_+ + 
\left(\frac{\ln(1-x)}{1-x}\right)_+ +{\cal O}(\epsilon)\right] 
\nonumber \\ 
&=&  C_F^2 S^{2\epsilon}x \left[ -\frac{1}{(-\epsilon)\epsilon} 
-\frac{1}{\epsilon}(\ln(1-x)-1) 
+\frac{1}{(-\epsilon)} \ln(1-x) +{\cal O}(\epsilon^0)\right]. 
\label{J1V1}
\end{eqnarray}

As only diagrams $(f)$ and $(g)$ have non-trivial IR singularity structure, 
we consider only the $(f)+(g)$ part of $J^{\prime(2)}$ and 
perform the IR subtraction there. Specifically, 
\begin{eqnarray}
J^{\prime(2)}_{(f+g)u}- J^{\prime(1)}\otimes V^{(1)} =  
x\left[ C_F^2 \frac{3}{2} \frac{1}{\epsilon}
+\frac{C_F C_A}{2} \frac{1}{2\epsilon}+{\cal O}(\epsilon^0) \right].   
\label{IRsubtract}
\end{eqnarray}
All IR singularities have been removed from $J^{\prime(2)}$ as dictated 
by its definition in Eq.~(\ref{Jprimedef}). 
The left over terms are simple UV poles.  
Now we are finally able to compute the contribution of the diagrams  
$(f)$ and $(g)$ to the jet anomalous dimension and to the UV renormalized 
jet function in $x$ space
\begin{equation}
\gamma^\prime_{J^\prime(f+g)}(\alpha_s) = -\frac{\alpha_s}{\pi}
\left(3 C_F+\frac{1}{2}C_A \right),
\label{gammaJ2fg} 
\end{equation}
and
\begin{equation}
J^{\prime(2)}_{(f+g)} = -\left(\frac{3}{2} C_F^2 +\frac{1}{4} C_F C_A \right) 
\ln(1-x) +{\rm reg.} 
\label{Jfgfin}
\end{equation}
It is true that we could have arrived at the above equation by just computing 
the UV simple poles of only the two particle cut graphs and converting them 
into $\ln(1-x)$ contributions. We, nevertheless, took the pain of 
calculating also the three particle cuts and the subtraction terms to show 
that indeed only simple UV poles survive in the final expression, thus 
showing that an anomalous dimension can be defined to sum the singular terms.

\subsection{The final  ${\cal O}(\alpha_s^2)$ result for $F_L$} 

First, the total contribution to $J^{\prime(2)}$ is obtained by summing 
the partial results in Eqs.~(\ref{J2a}, \ref{J2b}, \ref{J2c}, \ref{J2d}, 
\ref{J2e}, \ref{Jfgfin}). It is 
\begin{equation}
J^{\prime(2)} = \left[\frac{9}{4} C_F^2 -\frac{23}{12} C_F C_A 
-2\zeta(2) \left(C_F^2-\frac{C_F C_A}{2}\right) +\frac{1}{6} C_F N_f \right] 
\ln(1-x) +{\rm reg}. 
\label{J2tot} 
\end{equation}  
The leading soft double logarithm is generated by the UV renormalized 
$J^{\prime(1)}\otimes V^{(1)}$, Eq.~(\ref{J1V1}). 
It is 
\begin{equation}
(J^{\prime(1)}\otimes V^{(1)})(x) = \frac{1}{2} C_F^2 \ln^2 (1-x) +{\rm reg.} 
\label{J1V1fin} 
\end{equation}
We have organized the summation of the singular $\ln(1-x)$ terms 
in such a way that all fragmentation ${\cal O}(\ln(1-x))$ terms come from 
$J^{\prime(2)}$ and the soft ${\cal O}(\ln^2(1-x))$ term comes from 
$J^{\prime(1)}\otimes V^{(1)}$. 
At this stage we can determine the coefficient $\bar{C}_L^{(2)}(x,1)$ as
given 
by the $F_L$ factorization formula. It is 
\begin{equation}
\bar{C}_L^{(2)}(x,1) = (J^{\prime(1)}\otimes V^{(1)})(x)+J^{\prime(2)}(x),  
\label{cl2fin}
\end{equation}
which, through Eqs.~(\ref{J2tot}, \ref{J1V1fin}), 
exactly reproduces the singular part of the result from the full calculation 
in Eq.~(\ref{cl2}). 
We have thus shown explicitly how the factorization formalism captures the 
logarithmic enhancements to ${\cal O}(\alpha_s^2)$.    
Finally, the jet anomalous dimension $\gamma^\prime_{J^\prime}$ is
\begin{equation}
\gamma^\prime_{J^\prime}(\alpha_s) = \frac{\alpha_s}{\pi} 
\left[ \frac{9}{2}C_F -2 C_A - 4 \zeta(2) \left(C_F-\frac{C_A}{2}\right)
-\frac{1}{2}\beta_1 \right] +{\cal O}(\alpha_s^2). 
\label{gammaJtot}
\end{equation}
and the anomalous dimension $\gamma_{J^\prime}$ that does not include 
the running of the coupling is obtained from Eq.~(\ref{gammaprime}). It is 
\begin{equation}
\gamma_{J^\prime}(\alpha_s) = \frac{\alpha_s}{\pi} 
\left[ \frac{9}{2}C_F -2 C_A - 4 \zeta(2) \left(C_F-\frac{C_A}{2}\right)
\right] +{\cal O}(\alpha_s^2). 
\label{gammaJtotno}
\end{equation}

\section{Summary and conclusions}
\setcounter{equation}{0}

We have studied in detail the longitudinal structure function in deeply
inelastic  scattering near the elastic limit $x \rightarrow 1$ and have shown 
that it can be Sudakov factorized in terms of two jet and one soft functions. 
The soft function and the target jet function are the same as in $F_2$, 
whereas the current jet function is characteristic of $F_L$ and measures 
the transverse momentum of the struck parton.  
We have shown that the quasi-elastic logarithmic corrections from the 
fragmentation of the current jet can be resummed in moment space in terms 
of the anomalous dimension $\gamma_{J^\prime}$, 
which we calculated to ${\cal O}(\alpha_s)$.  

We emphasize that our analysis is at the leading twist level but goes beyond 
the light-cone expansion. 
It resums terms that are of order $\ln^kN/N, k>0$. 
This approach can be extended to the case of $F_2$, where such terms are 
present, but subleading relative to the order $\ln^kN,\  k \ge0$ terms. 
It is quite remarkable that terms that are power suppressed 
in $N$ are amenable to a similar systematic analysis as the leading ones, 
just through the introduction of a single new function, the jet $J^\prime$. 
This leads us to the conjecture that, indeed, all power 
in $N$ suppressed terms can be resummed in quasi-elastic processes,
via the introduction of further jet functions defined by operators with 
higher powers of the transverse derivative. 
In Ref.~\cite{ASS1} a systematic procedure has been presented for contructing 
these higher dimensional jet functions, and the present analysis is really 
a concrete example of that procedure.

Returning to $F_L$, we note also that the logarithmic enhancements from 
the final state fragmentation are delayed by one order of $\alpha_s$ 
relative to the ones in $F_2$. 
As these enhancements are summed to all orders, 
part of their effect (the beta function term in Eq.~(\ref{flsol})) is 
to change the overall normalization by shifting the scale 
of $\alpha_s$ in the preexponent factor from $Q^2$ to $Q^2/N$. 
Thus, in the region $x \rightarrow 1$ higher order corrections are 
parametrically as important as lower order ones. This has implications 
for obtaining the power corrections to $F_L$, using for instance renormalon 
methods, since the power corrections come predominantly from this near elastic
region. 
This enhancement, which is characteristic of $F_L$, can be thought of 
as an indication from perturbation theory of the large power corrections 
that this observable is known to receive \cite{renorm}. 
These and applications of the formalism to other inclusive processes are
currently under investigation.
 
\bigskip

\noindent
{\it Acknowledgement}:  We are very grateful to George Sterman for dicussions and suggestions. This
work is supported in part by  the U.S. Department of Energy. 

\bigskip 

%\appendix
\section*{Appendix}

\noindent
$\bullet$ Three useful integrals with both quadratic and linear denominators, 
valid at $p^2=0$:
\begin{equation}
\int \frac{d^D l}{(4 \pi)^D} \frac{1}{(l-\bar{p})^2 (l-k)^2 p\cdot l}
= \frac{-i}{(4 \pi)^2} \frac{1}{\epsilon} \frac{1}{p\cdot (\bar{p}-k)}
\ln \left(\frac{p\cdot k}{p\cdot \bar{p}} \right)
+{\cal O}(\epsilon^0) \, .
\label{cool1}
\end{equation}
\begin{equation}
\int \frac{d^D l}{(2 \pi)^D} 
\frac{l^\mu}{(l-\bar{p})^2 \, l^2 \, p \cdot l} = \frac{i}{(4 \pi)^2} 
\frac{1}{\epsilon} \frac{\bar{p}^\mu}{p\cdot \bar{p}}
 +{\cal O}(\epsilon^0) \, . 
\label{cool2}
\end{equation}
\begin{equation}
\int \frac{d^D l}{(4 \pi)^D} \frac{l^\mu l^\nu}
{(l-\bar{p})^2 \, (l-k)^2 \, l^2 \, p\cdot l} = 
\frac{-i}{(4 \pi)^2} \frac{1}{2 \epsilon}
\frac{ \eta_{\mu \nu} }{p \cdot (\bar{p}-k)} 
\ln \frac{p \cdot k}{p \cdot \bar{p}} +{\cal O}(\epsilon^0) \, .
\label{cool3}
\end{equation}

\medskip

\noindent
$\bullet$ The three massless particle phase space: \\
The three particle phase space cannot be written conveniently in terms of the 
light cone momenta of the outgoing particles alone. In this case we prefer
the 
parametrization in terms of Lorenz invariants defined in the rest frame 
of two of the three particles, say $k_2$ and $k_3$. 
Specifically, 
\begin{eqnarray}
\int dPS_3(k_1, k_2,k_3) &=& \frac{1}{(4 \pi)^D} \frac{1}{\Gamma(D-3)}
\left(\frac{(1-x)Q^2}{x}\right)^{D-3} 
\int_0^\pi d\theta \, \sin^{D-3}\theta \int_0^\pi d\phi \, \sin^{D-4}\phi   
\nonumber \\
&\quad&\times 
\int_0^1 dy_1 \, y_1^{D/2-2}(1-y_1)^{D-3} 
\int_0^1 dz \, z_1^{D/2-2} (1-z_1)^{D/2-2}, 
\label{PS3}
\end{eqnarray} 
where $\theta$ and $\phi$ are the polar and azimuthal angles of $\vec{k_1}$
with respect to the $\vec{p}$ direction. In the $k_2$, $k_3$ CM frame, 
$\vec{p}$ defines the direction of the light cone, which is different from
the 
one defined in Eq.~(\ref{vdef}). The definitions of $y_1$ and $z$ are 
\begin{equation}
y_1= \frac{p\cdot k_1}{p \cdot\bar{q}}, \ \ \ 
z_1=\frac{(1-x)-y_1- (q\cdot k_1)/(p\cdot \bar{q})}{(1-x)(1-y_1)}.
\label{yzdef}
\end{equation}
It follows that 
\begin{equation} 
\frac{p\cdot k_2}{p \cdot \bar{q}}=(1-y_1)\left(\frac{1-
\cos\theta}{2}\right), 
\ \ 
\frac{p\cdot k_3}{p \cdot \bar{q}}=(1-y_1)\left(\frac{1+\cos\theta}{2}\right),
\ \
\frac{k_2\cdot k_3}{p\cdot \bar{q}}=(1-x)(1-y_1)z_1.
\label{kkin}
\end{equation}
For azimuthically symmetric integrands the $d\phi$ integration can be readily
performed and in terms of the variable 
\begin{equation}
\omega=\frac{1-\cos\theta}{2} 
\label{changevars}
\end{equation}
the phase space becomes
\begin{eqnarray}
\int dPS_3(k_1, k_2,k_3) &=& \frac{1}{2(4\pi)^3} 
\left(\frac{(1-x) Q^2}{x}\right) 
\left(\frac{4\pi x}{(1-x)Q^2}\right)^{2\epsilon} 
\frac{1}{\Gamma^2(1-\epsilon)} \int_0^1d\omega \,\omega^{-\epsilon}
(1-\omega)^{-\epsilon} 
\nonumber \\
&\ & \times \int_0^1dy_1\,y_1^{-\epsilon}(1-y_1)^{1-2\epsilon} 
\int_0^1dz_1\, z_1^{-\epsilon} (1-z_1)^{-\epsilon}. 
\label{PS3fin}
\end{eqnarray}
Since all three particles are massless, the phase space is symmetric in their
variables. So the phase space integrals in the $k_1$, $k_3$ CM system are
obtained 
from the above expressions by permutation of the particle indices. 

\medskip

\noindent
$\bullet$ Finally, certain useful convolution lemmata:
\begin{equation}
h(x) \otimes \delta(1-x) = h(x).
\label{+1}
\end{equation}
\begin{equation}
x\otimes \left(\frac{1}{1-x}\right)_+=
x \left(\ln(1-x)-\ln x+\frac{1}{x}-1\right). 
\label{+2}
\end{equation}
\begin{equation}
x\ln(1-x)\otimes\left(\frac{1}{1-x}\right)_+ = 
x\ln^2(1-x) +(1-x)\ln(1-x)-(1-x)-x\zeta(2). 
\label{+3}
\end{equation}
\begin{equation}
x\otimes \left(\frac{\ln(1-x)}{1-x}\right)_+ = 
\frac{1}{2}x\ln^2(1-x) +\frac{(1-x)}{x}\ln(1-x) +x\ln x.
\label{+4}
\end{equation}
\begin{equation}
x\otimes(1+x^2)\left(\frac{\ln(1-x)}{1-x}\right)_+ = 
x\ln^2(1-x)+{\rm reg.}
\label{+5}
\end{equation}

\begin{figure}
\centerline{\epsfig{file=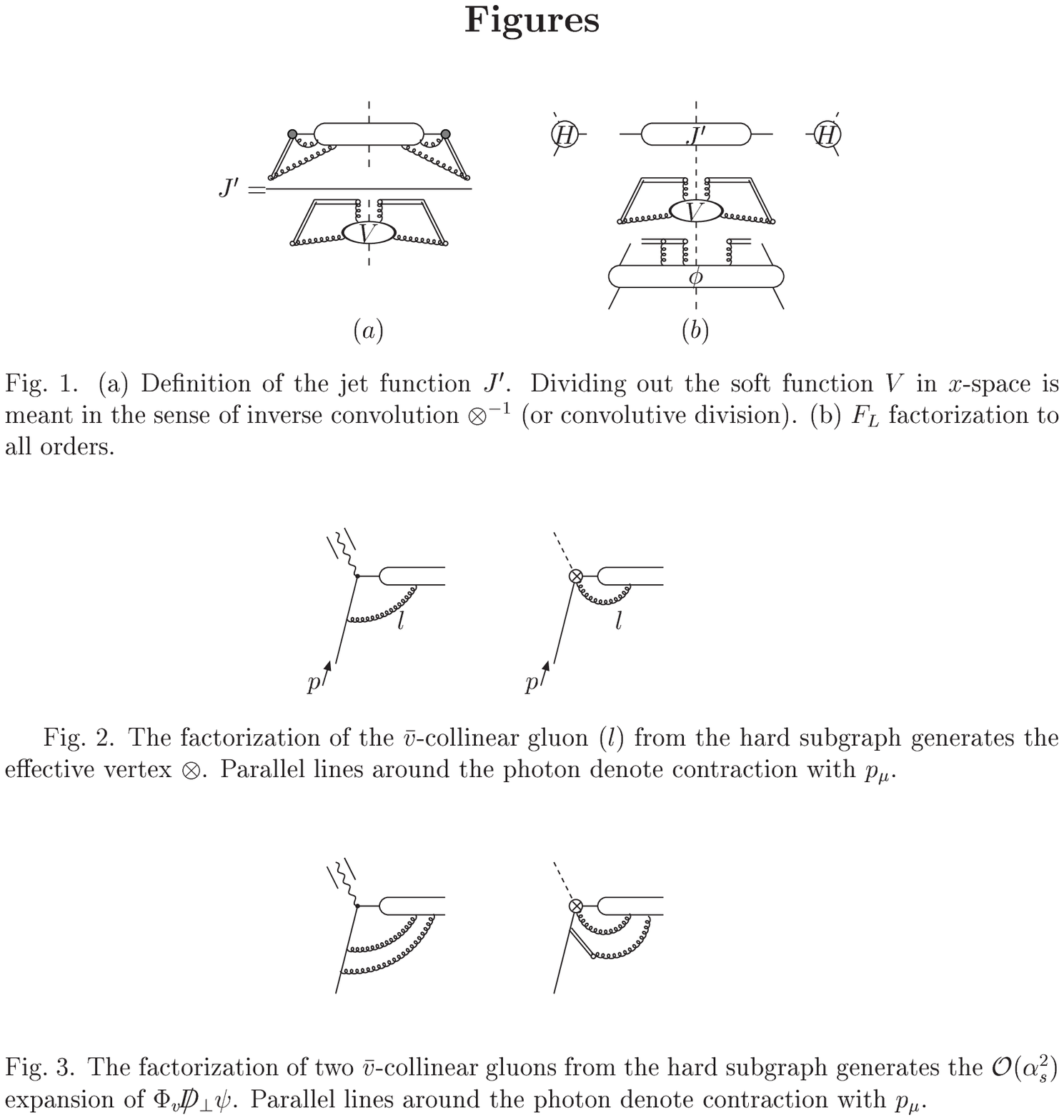,scale=1}}
\end{figure}

\begin{figure}
\centerline{\epsfig{file=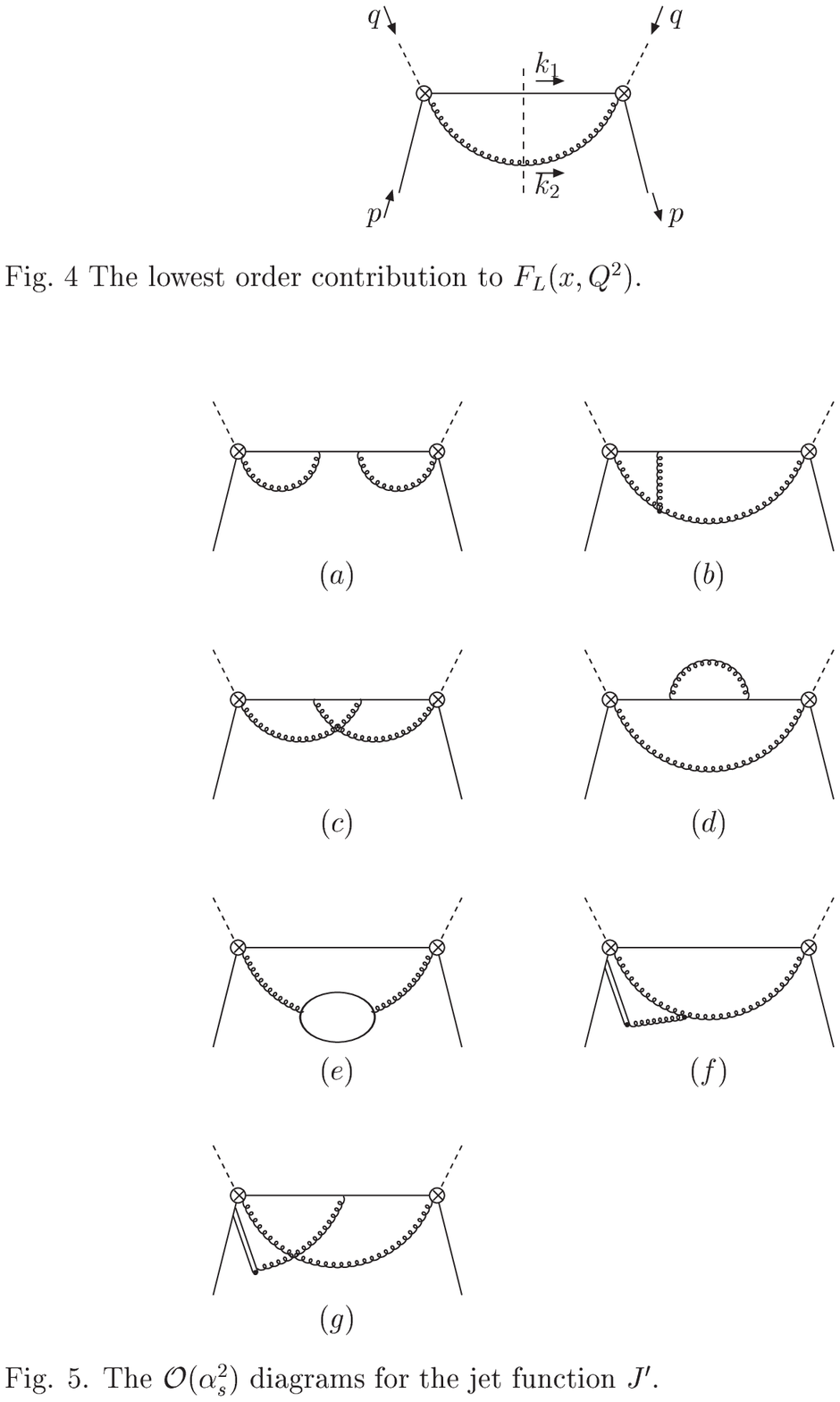,scale=1}}
\end{figure}

\begin{figure}
\centerline{\epsfig{file=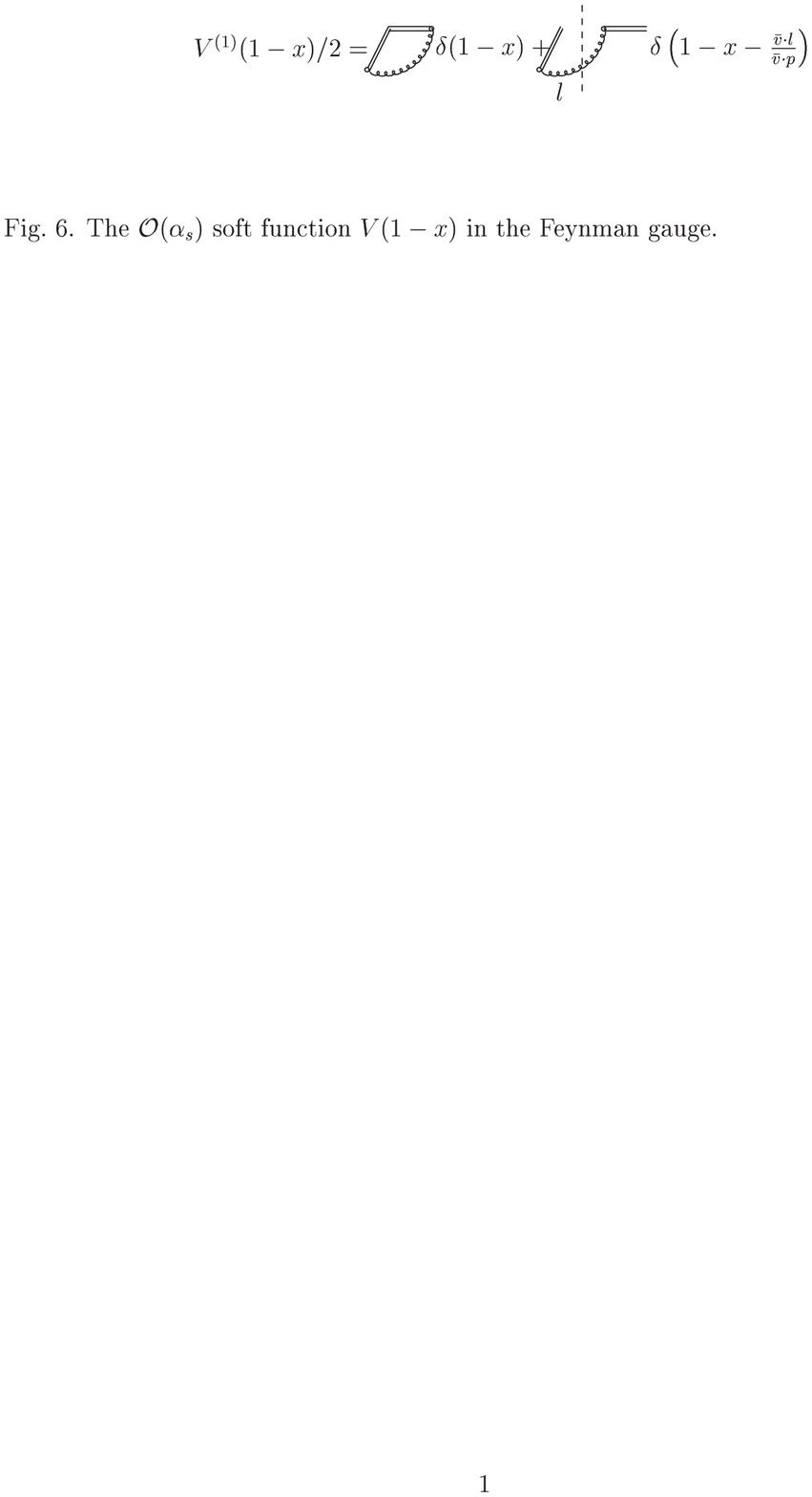,scale=1}}
\end{figure}

\end{document}